\documentclass[fleqn,usenatbib]{mnras}

\usepackage{newtxtext,newtxmath}
\usepackage[font=footnotesize]{caption}
\usepackage{caption}
\usepackage[T1]{fontenc}

\DeclareRobustCommand{\VAN}[3]{#2}
\let\VANthebibliography\thebibliography
\def\thebibliography{\DeclareRobustCommand{\VAN}[3]{##3}\VANthebibliography}

\usepackage{graphicx}	
\usepackage{amsmath}	

\title[Kron 3: A cluster with an extended red clump in the UV]{UVIT-{\it HST}-{\it Gaia}-VISTA study of KRON 3 in the Small Magellanic Cloud: A cluster with an extended red clump in UV }

\author[P. K. Nayak et al.]{
P. K. Nayak,$^{1,2}$\thanks{E-mail: nayakphy@gmail.com}
A. Subramaniam,$^{1}$
S. Subramanian,$^{1}$
S. Sahu,$^{1}$
C. Mondal,$^{1}$ 
\newauthor Maria-Rosa L. Cioni,$^{3}$
Cameron P. M. Bell,$^{3}$
A. Bandyopadhyay$^{4}$
and Chul Chung$^{5}$
\\ \\
$^{1}$Indian Institute of Astrophysics, 2nd Block, Koramangala, Bangalore, 560034, India \\
$^{2}$Tata Institute of Fundamental Research, Homi Bhabha Road, Navy Nagar, Colaba, Mumbai 400005, India \\
$^{3}$Leibniz Institute for Astrophysics Potsdam (AIP), An der Sternwarte 16, D-14482 Potsdam, Germany \\
$^{4}$Aryabhatta Research Institute of Observational Sciences, Uttarakhand 263002, India \\
$^{5}$ Department of Astronomy \& Center for Galaxy Evolution Research, Yonsei University, Seoul 03722, Republic of Korea
}
\date{Accepted 2021 February 3. Received 2021 January 18; in original form 2019 December 22}

\begin{document}
\label{firstpage}
\pagerange{\pageref{firstpage}--\pageref{lastpage}}
\maketitle

\begin{abstract} 
We have demonstrated the advantage of combining multi-wavelength observations, from the ultraviolet (UV) to near-infrared, to study Kron 3, a massive star cluster in the Small Magellanic Cloud.  
We have estimated the radius of the cluster Kron 3 to be 2.$'$0 and for the first time, we report the identification of NUV-bright red clump (RC) stars and the extension of the RC in colour and magnitude in the NUV vs (NUV$-$optical) colour-magnitude diagram (CMD).
We found that extension of the RC is an intrinsic property of the cluster and it is not due to contamination of field stars or differential reddening across the field.  
We studied the spectral energy distribution of the RC stars, and estimated a small range in temperature $\sim$  5000 - 5500 K,  luminosity $\sim$ 60 - 90 L$_\odot$ and radius $\sim$ 8.0 - 11.0 R$_\odot$ supporting their RC nature. The range of UV magnitudes amongst the RC stars ($\sim$23.3 to 24.8 mag) is likely caused by the combined effects of variable mass loss, variation in initial helium abundance (Y$_{ini}$=0.23 to 0.28),  and a small variation in age (6.5-7.5 Gyr) and metallicity ([Fe/H]=$-$1.5 to $-$1.3). 
Spectroscopic follow-up observations of RC stars in Kron 3 are necessary to confirm the cause of the extended RC. 

\end{abstract}




\section{Introduction}

Star clusters were known to be the best examples of coeval systems and thought to follow simple stellar population models. Recent Hubble Space Telescope (HST) observations of Galactic globular clusters (GCs) and massive star clusters in the Magellanic Clouds (MCs), however, have changed our understanding of cluster formation. 
It is found that intermediate age (mostly younger than 2 Gyr) massive ($\sim$ a few times 10$^4$ M$_\odot$) clusters in the MCs show an extended main sequence turn-off (eMSTO), which can not be explained through photometric errors or binarity present in the clusters \citep{mackey2008, milone2008, milone2009}.  
Intrinsic age spreads and the effects of stellar rotation are suggested to explain the presence of such eMSTOs. 
Recent studies of Galactic open clusters (OCs) using the Gaia Data Release (DR) 2
\citep{bastian2018,cordoni2018,marino2018,lim2019,sun2019,li2019,goss2019} also show extended features in both main sequence (MS) and MSTO in the colour-magnitude diagrams (CMDs) and it is suggested that the distribution of the projected rotational velocity of stars play a major role in the creation of these extended features. 
The Galactic GCs (older than 10 Gyr) and relatively older (more than 2Gyr old) massive clusters in the MCs show multiple evolutionary sequence in the MS, sub-giant and red giant branch (SGB \& RGB) \citep{piotto2007, anderson2009, piotto2009}. Multi-modal populations in age and/or enrichment in chemical abundance (i.e. spread in He, C, N, O, Na etc.) have been proposed to explain the observed split in different evolutionary sequences. 
It is well known that there is a paucity of clusters in the age range 4-9 Gyr in the Large Magellanic Cloud (LMC) \citep{vanden1991} as well as in our Galaxy, whereas, the Small Magellanic Cloud (SMC) hosts a relatively large number of rich clusters in the above age range. \cite{neider2017a,neider2017b} studied four clusters in the SMC in the above age ranges, which are similarly massive but significantly younger ($\sim$6 Gyr old) to Galactic GCs, show an eMSTO as well as a split in the RGB. The presence of a multiple stellar population with a variation in the N abundance causes the splitting in the RGB. The second population of RGB stars is enhanced in N and appears redder in the CMD.
Kron 3 is a massive intermediate-age ($\sim$6.5 Gyr old; \citealp{glatt2008}) star cluster located to the west of the main body of the SMC. 
The study of Kron 3 will add to the analysis of clusters of this type as well as provide additional constraints to help us better understand multiple stellar population in massive star clusters.

\cite{gasco1966} presented the first CMD of Kron 3 and reported it to be an intermediate-age ($\sim$2 Gyr) cluster. 
\cite{gas1980} estimated the age of the cluster to be 3 Gyr with deeper photometric data (limiting magnitude $V$=21 mag). 
Comparing its CMD with theoretical stellar evolutionary models, \cite{hodge1982} reported a metallicity and age of Kron 3 of [Fe/H] = $-$1.3 ($\pm$0.3) dex and 1 Gyr ($\pm$0.4), respectively.

Early studies of Kron 3, however, were biased as a result of shallow photometric data that did not reach the MS of the cluster. \cite{rich1984} detected the MSTO for the first time (limiting magnitude $R$=23 mag), using observations from the CTIO 4m telescope and found that Kron 3 has an age range of 5-8 Gyr. The authors also mentioned that the uncertainty in the distance modulus (DM) prevented a more precise estimate of the cluster age.
The authors estimated the radius of the cluster to be $\sim$2$'$.4 (42 pc) using the distribution of star counts around the cluster center. 
By fitting isochrones \citep{vanden1985} to the  CMDs, \cite{alca1996} showed that the age of Kron 3 ranges from 8-10 Gyr with a  DM value of 18.75 mag. They suggested that estimates of a higher DM could be the reason for getting younger ages in previous studies. 
They questioned the previous estimation of radius by \cite{rich1984}. \cite{alca1996} showed that stars located in a field 2.$'$0 (35 pc) from the cluster center could reproduce cluster RGB and red clump (RC) in the CMD, along with a rich population of MSTO stars. Therefore, they claimed that the radius of the cluster is larger ($\sim$ 6$'$ or 105 pc) than previously estimated ($\sim$ 2.$'$4 or 42 pc) by \cite{rich1984}. On the other hand, the recent study by \cite{bica2008} suggested that the radius of the cluster as 1.$'$7 (30 pc).

With the help of archival HST Wide Field and Planetary Camera 2 (WFPC2) observations in F450W ($B$) and F555W ($V$) bands, \cite{migh1998} estimated the age of the cluster as 4.7($\pm$0.6)Gyr. 
The most recent study by \cite{glatt2008} using HST Advanced Camera for Surveys (ACS) data suggests the age to be 6.5 Gyr, with a metallicity of $Z$=0.001. 
Hence, there is a range of ages (1-10 Gyr) and radii ($\sim$ 30-105 pc) suggested for this cluster.

The metallicity of this cluster is also a source of debate. \cite{gas1980} estimated the metallicity to be [Fe/H] = $-$0.6 dex  using spectra (range 3700-6300 \AA) acquired from the Anglo-Australian Telescope (AAT), whereas the broad-band Canterna photometry of individual giants resulted in a metallicity of [Fe/H] = $-$1.5 dex. 
Studies by \cite{hodge1982}, \cite{rich1984}, \cite{alca1996}, \cite{glatt2008} suggest that the metallicity ($Z$) of the cluster is 0.001 ([Fe/H]=$-$1.3 dex), whereas \cite{dias2010} reported the metallicity of the cluster as Z=0.0002 by comparing the integrated spectra of the cluster with single stellar population models, suggesting it to be a very metal poor cluster.  
\cite{alca1996} noticed an increase in the RGB slope in the CMD with increasing radius from the cluster and proposed that a wide range in metallicity could be the reason for this behaviour.  
A recent study by \cite{holly2018} confirmed the spectroscopic evidence of nitrogen-enhanced stars among the RGB stars. The presence of a large range of elemental abundance indicates that Kron 3 hosts multiple stellar populations. As the ultra-violet (UV) regime is more sensitive to metallicity, due to the presence of a large number of absorption lines, it is important to study the UV properties of this cluster, which have not yet been explored.

In this study, we focus on the following aspects: (a) an estimate of the radius of the cluster and (b) a study of the age and metallicity spreads using multi-wavelength data.  
In order to achieve the above aims, it is important to have a large and contiguous areal coverage as the HST studies have primarily focused on the central region, whereas the outer regions of the cluster have only been studied using ground-based data. It is important to have a large wavelength coverage, including the UV for the cluster as well as the field. We combine UV observations of Kron 3 taken with the Ultra-Violet Imaging Telescope (UVIT) with HST (for the central regions of the cluster), {\it Gaia} and the Visual and Infrared Survey Telescope for Astronomy (VISTA) observations (for the outer regions of the cluster). The superior resolution of UVIT [$\sim$1.$''$5, three times better than that from the Galaxy Evolution Explorer (GALEX)] in the near-UV and the large field-of-view (28$'$ diameter) are the main advantages of this study.

The remaining sections of this paper are arranged as follows. In Sect. 2, we discuss the NUV data used for this study and use these to determine a new estimate of the cluster radius. In Sect. 3, we present UV-Optical CMDs using UVIT and HST that demonstrate the extended nature of the RC in the NUV. In the Sect. 4, we discuss the effect of photometric zero points variations in the UV-optical CMDs. In Sect. 5, we present UV-Optical CMDs using UVIT and HST and have discussed the effect of field star contamination in these CMDs. Sect. 6 provides the analysis to find the possible reasons behind the extended RC in NUV. We discuss the results of this paper in Sect. 7 and summarise with our conclusions in Sect. 8.

\section{UVIT photometry and cluster radius}

\begin{figure}
\centering
\includegraphics[width=\columnwidth]{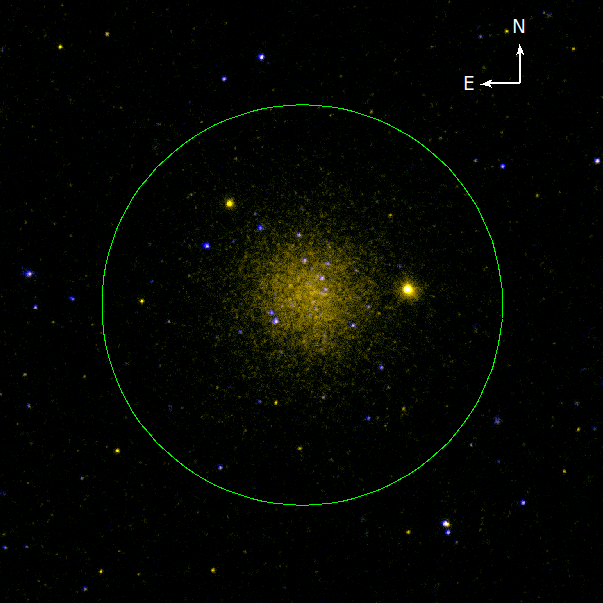}
\caption{Colour composite $6' \times 6'$ image of the Kron 3 cluster. Yellow and blue colours correspond to stars detected in NUV and FUV band, respectively. The green circle indicates 2$'$ radius around the cluster.}
\label{image}
\end{figure}

The observations of Kron 3 were carried out with the UVIT telescope in far-UV (FUV) and near-UV (NUV) bands as part of a Guaranteed-Time proposal (G08) on 25th March 2018. Kron 3 was observed in one FUV (F148W : 125-175 nm) and one NUV (N242W : 203-281 nm) filter. The total exposure time was 7194 seconds. The observations were completed in multiple orbits. We applied corrections for spacecraft drift, flat-field and distortion,  using the software CCDLAB \citep{postma2017} and created images for each orbit. Then, the orbit-wise images were co-aligned and combined to generate science ready images. 
The science ready images were created for an area of 4K$\times$4K in size with a scale of 0.$''$4125/pixel. The calibration of the instrument can be found in \cite{tandon2017a,tandon2020}. The details of the telescope and instrument are available in \cite{tandon2017b} and \cite{subra2016}.
Figure \ref{image} shows a colour composite image of the cluster, Kron 3. Stars detected in NUV are shown in yellow, while the FUV detected stars are shown in blue.

We used the DAOPHOT/IRAF tasks and packages \citep{stetson1987} to carry out the photometry. The FWHM of the model PSF for FUV and NUV filters are 1.$''$24 and 1.$''$03, respectively. To detect the sources, we used threshold as six times the average background count. Due to crowding in the central regions of the cluster, we have performed PSF photometry to the detected stars. First, a model PSF was generated using isolated stars for both the science ready FUV and NUV images. Then the model PSF has been fitted to all the detected stars to select the stars with good PSF values with their PSF magnitude. 
We applied  aperture corrections and saturation corrections to PSF magnitudes and calculated the final magnitudes of the detected stars in the corresponding bands by adding zero point magnitudes. The values of zero point magnitudes have been taken from \citet{tandon2020}.

\begin{figure}
\centering
\includegraphics[width=\columnwidth]{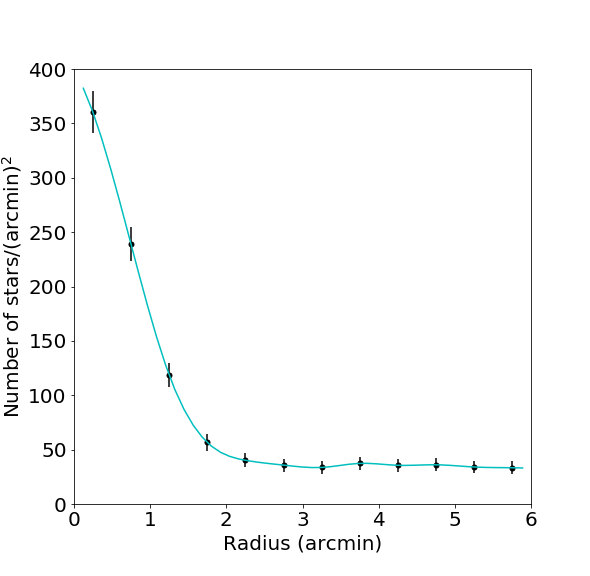}
\caption{ Radial distribution of stellar density at 0.$'$5 intervals estimated from NUV data. The profile suggests a cluster radius of $\sim$2.0$'$.}
\label{king_uvit}
\end{figure}

\begin{figure}
\centering
\includegraphics[width=\columnwidth]{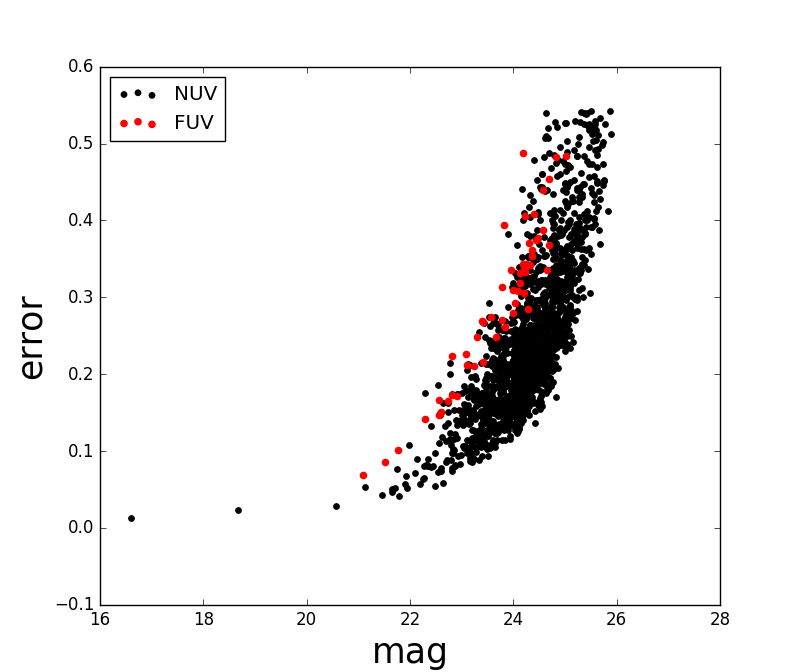}
\caption{Photometric errors of stars within the cluster radius (2.0$'$) as a function of magnitude in both NUV and FUV bands.}
\label{err}
\end{figure}

Figure \ref{image} shows that there are only a few FUV detections in the central parts of Kron 3. Therefore, we used NUV data to estimate the cluster radius. We counted the number of stars present in bins of 0.$'$5 in radius around the cluster center and normalized with respect to the area of the corresponding annuli to determine the density of stars in each annulus. We plotted the number density of stars as a function of radius in Figure \ref{king_uvit} and fitted it with a cubic spline function. The fitted curve suggests that the density of stars within the cluster region merges with that of field stars at $\sim$2.$'$0. 
For the remainder of this study, we will adopt a cluster radius of 2$'$.0. Figure \ref{err} shows the photometric error of stars within the cluster radius as a function of PSF magnitudes for NUV (black points) and FUV (red points) bands. The figure suggests that the photometric errors are $\le$ 0.3 mag in both bands at magnitude $\sim$24.

\subsection{Artificial Star test}

\begin{figure}
\centering
\includegraphics[width=\columnwidth]{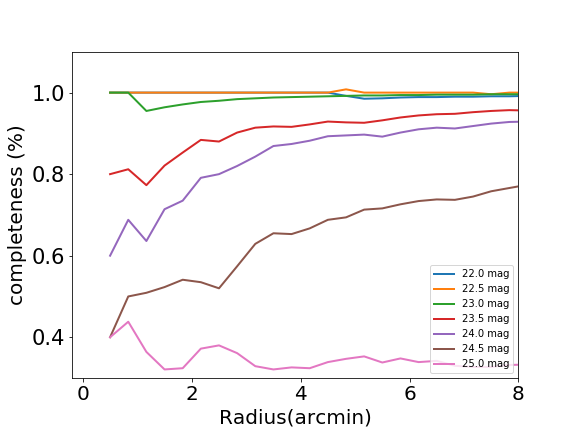}
\caption{Completeness as a function of radius and V magnitudes.}
\label{complete}
\end{figure}

Artificial-star (AS) tests provide a crucial role to determine the completeness level of the data used in this study. We have added artificial stars to original image with a spatial density distribution of stars similar to the distribution of original stars.  
For each iteration, the number of inserted artificial stars were 20\% of detected stars keeping the magnitude same for all stars and the positions of the stars were generated randomly. 
The detected NUV stars have a magnitude range 22.0 to 25.0 magnitude and we have divided the range in 9 bins to perform the AS test for the corresponding magnitude. To recover added stars from the image, we followed similar procedures as used for the real stars. 
We have considered a star to be recovered when it has a magnitude difference less than 0.75 mag and spatial difference less than 0.5 pixel. The completeness is measured as a ratio of recovered stars to the added stars for a fixed magnitude. Figure \ref{complete} shows the completeness of NUV data as a function of cluster radius for different NUV magnitudes. We found that the data reached $\sim$50\% of its completeness level at 24.5 mag at the densest or core region ($<$0.5$'$) of the cluster. 

\section{colour-magnitude diagrams}

\subsection{UV-Optical CMD using UVIT and HST data}

Only 54 stars were detected in the FUV band compared to 1623 stars in the NUV within a cluster radius of 2$'$.0, making it hard to construct a FUV$-$NUV CMD to study the properties of the cluster. 
On the other hand, we can combine NUV data with data obtained at optical and near-infrared (near-IR) wavelegths. 
This will give us a broader colour range for the various evolutionary sequences of the cluster, particularly near the MSTO region which is used to estimate the fundamental parameters of the cluster (e.g. age).  
We used the publicly available optical photometric (DAOPHOT) data in F555W ($\sim V$) and F814W ($\sim I$) filters obtained from the Hubble Legacy Archive (HLA), that covers the central region of the cluster. We considered stars brighter than 22.5 mag in V while cross-matching with the NUV data because stars fainter than this are not detected in the NUV image due to the sensitivity limit of UVIT.   
We have compared the isochrone with the general sensitivity of UVIT to put the limit in F555W.  
Not considering stars fainter than 22.5 magnitude in V also helped to reduce the crowding in the central part of the cluster, making it easy to cross-match with the NUV data. We cross matched the NUV detections with HLA data by considering a maximum separation of 1.$''$0. 
In our analysis we excluded those common detections where we found multiple HLA detections within 1.$''$0 radius of a given NUV stars. There are 165 such NUV detections. The number of multiple detections increases rapidly for separation larger than 1.$''$0. Using the above mentioned criteria, we found 749 detections in common between both data sets.

\begin{figure*}
\centering
\includegraphics[width=1.9\columnwidth]{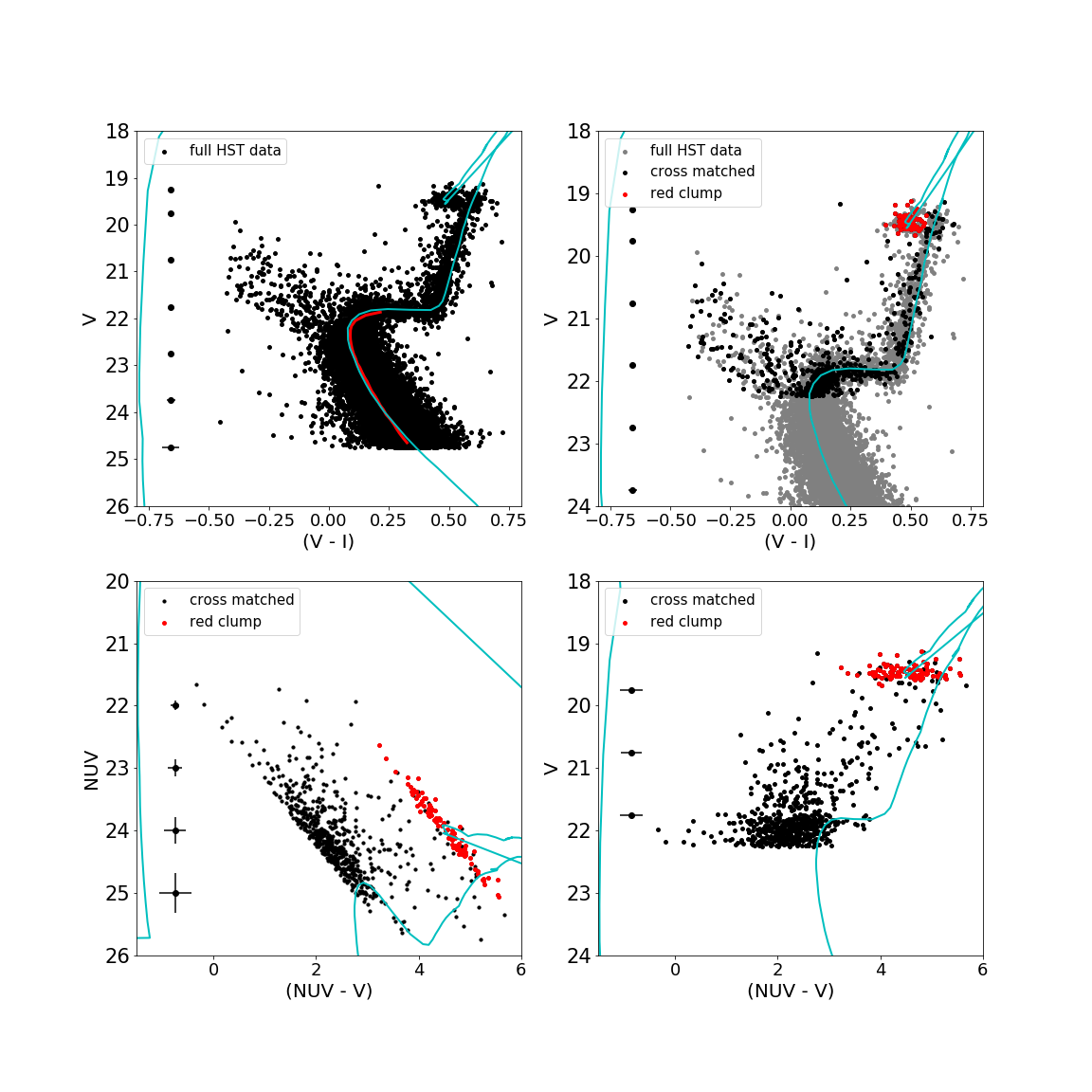}
\caption{(top-left) CMDs of HLA stars within the central 2’ of Kron 3. Red line indicated the fiducial line of MS stars. (top-right) CMDs of stars cross-matched between HLA and NUV data (black) over-plotted onto the full HLA sample (grey) with RC stars marked in red. The cross-matched stars are shown in the NUV vs (NUV-V) CMD (bottom-left) and V vs (NUV-V) CMD (bottom-right) where the distribution of RC stars appears stretched. In all panels an isochrone (cyan) corresponding to an age of 7 Gyr and a metallicity Z=0.001 is overlaid onto the CMDs after correcting for extinction and DM.}
\label{cmd_hst}
\end{figure*}

In the top-left panel of Figure \ref{cmd_hst}, we plotted $V$ vs ($V-I$) CMD of the full HLA data of Kron 3. We noticed a large spread in the MS and MSTO region. We also over-plotted a Padova isochrone (\citet{marigo2007,marigo2008} ; cyan line) of age 7 Gyr and metallicity Z=0.001, after correcting for reddening using $E(V-I)$ = 0.033 mag and a DM of 18.8 mag. We took the age, reddening, DM and metallicity values from \cite{glatt2008}. To generate the isochrone, we have used the Flexible Stellar Population Synthesis (FSPS) model \citep{conroy2009,conroy2010} and convolved the Padova models with the effective area curves of different filters to obtain the model magnitudes. 
The aforementioned reddening and DM are used throughout this study to fit the CMDs. 
In the top right panel we highlighted cross-matched stars and in particular, the RC stars. We plotted also an isochrone.  
The bottom right panel shows $V$ vs ($NUV-V$) CMD of only cross matched stars (black points) and RC stars, which are highlighted in red. We fitted an isochrone after correcting for reddening, extinction and DM. We can see that the RC is no longer a clump in the UV-optical CMD, instead it appears stretched along the colour axis. The extension of RC stars towards both bluer and redder colours indicates that the RC stars span a range of NUV magnitudes. Furthermore, the SGB, RGB and a part of MS stars are not fitted well by a single isochrone, while they are fitted well in the optical CMD. The spread in the MSTO region noticed in the $V$ vs ($V-I$) CMD becomes even larger in the V vs ($NUV-V$) CMD.  We find that a few SGB, RGB and MS stars get brighter in NUV and appear bluer in ($NUV-V$). In the bottom-left panel we plotted HLA-NUV cross-matched stars, highlighting RC stars, in the NUV vs ($NUV-V$) CMD. 
This plot clearly shows that RC stars extend more than two magnitudes in both colour and magnitude.  We also notice that UVIT reaches its detection limit near the MSTO region, and the SGB as well as part of the RGB become fainter than the MSTO in the NUV band. Photometric errors are also relatively large near the detection limit. The bottom of the RGB even exceeds the detection limit of UVIT. Therefore, the stars near the bottom of the RGB in the NUV-optical CMDs not fitted by the isochrone could be due to a poor detection in NUV. Henceforth, we focused our study on the evolved stellar population, mainly RC stars which are detected in NUV with less uncertainties. 
The extension of the RC shall not be due to photometric errors.

The observed extension in the RC could be an effect of uncertainties in the photometric zero points. 
The variation can occur due to differential reddening in the cluster region and/or to photometric inaccuracies. Poor sky determination or not accounting for the spatial variation of the PSF to generate the PSF model causes inaccuracies in photometry \citep{milone2012}. Therefore, we need to check whether differential reddening or PSF variations across the field can cause the observed extension in RC.


\section{Photometric zero point variations}

Typically, it is assumed that the foreground dust distribution is uniform across a given cluster. In some cases, however, it has been found to be patchy, resulting in reddening variations across the cluster and leading to an artificial broadening in the stellar sequences on the CMDs \citep{milone2012}. Another source of non-intrinsic broadening can manifest due to PSF variations across the CCD, leading to slight shifts in the photometric zero point (see Anderson et al.2008 for details). 
To unambiguously determine the presence of multiple populations within a given cluster, one must first account the main sources of non-intrinsic scatter/broadening within the photometry. In this Section, we investigate the effects of both differential reddening across the cluster as well as PSF variations in the photometric zero points to assess if they have a significant impact on the inferred spread of the RC.

\subsection{Differential reddening}

To check for differential reddening, we adopted the method by \citet{milone2012}, in which they used MS stars as a tool to estimate the effect of reddening. As we have only a few detections in the FUV to make a UV CMD and the UVIT reaches its detection limit near the turn-off of Kron 3, we instead used HST data to calculate the reddening levels across the cluster and converted these to the corresponding values in the NUV band to de-redden the NUV-optical CMD.

\begin{figure*}
\centering
\includegraphics[width=1.9\columnwidth]{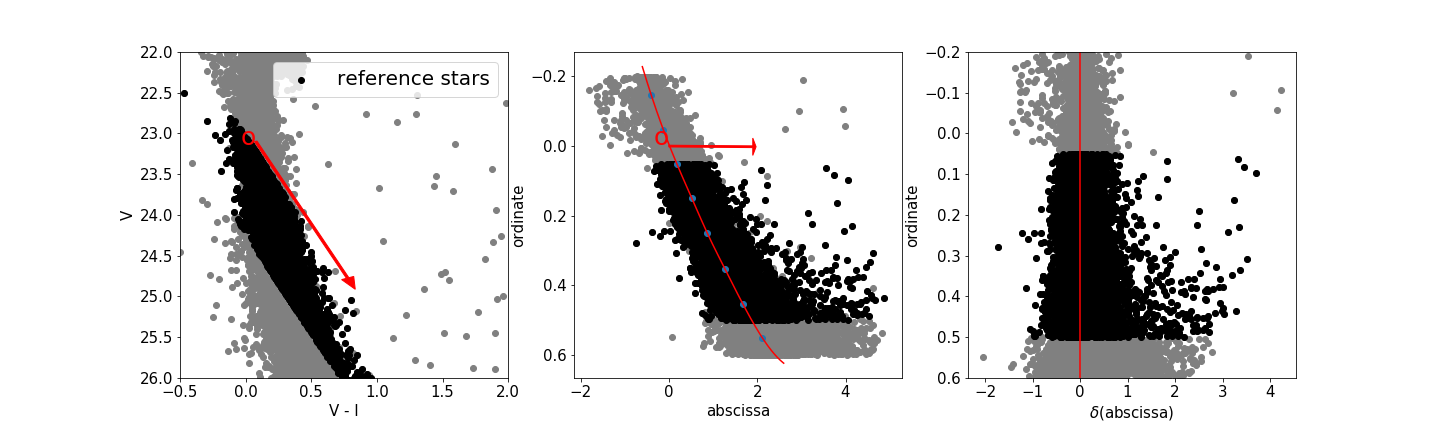}
\caption{(left) CMD of HLA stars within the central 2’ of Kron 3. The red arrow indicates the reddening vector, also defined as abscissa axis and parallel to the arrow is defined as ordinate axis. (middle panel) distribution of stars in ordinate vs abscissa reference frame. Red line is the MS fiducial line in reference frame. Black points are reference stars used to determine differential reddening. (right panel) Distribution of stars with respect to fiducial line. }
\label{cmd_rot}
\end{figure*}

We briefly summarise the method below, but refer the reader to Sec. 3.1 of \citet{milone2012} for more details. 
First we define an arbitrary point near the MSTO (o) and then with respect to that point rotate the CMD counter-clockwise by an angle theta=$\arctan$ [$A_V/(A_V - A_I)$] so that the abscissa became parallel to the reddening vector. $A_V$ and $A_I$ are nothing but the absorption coefficients in the V and I bands corresponding to the average reddening of the cluster. 
The reason for rotating the CMD is that it is much more intuitive to determine a reddening difference on the horizontal axis rather than along the oblique reddening vector. 
Now “abscissa” and “ordinate” will be indicated as the abscissa and ordinate of the rotated reference frame, respectively. 
The left panel of Figure \ref{cmd_rot} shows the highlighted (black points) region of MS, used as reference MS stars to determine the effect of differential reddening. The red arrow indicates the reddening vector, also defined as direction of abscissa axis of the rotated CMD. The middle panel shows the rotated CMD with respect to the point "o".


\begin{figure*}
\centering
\includegraphics[width=1.9\columnwidth]{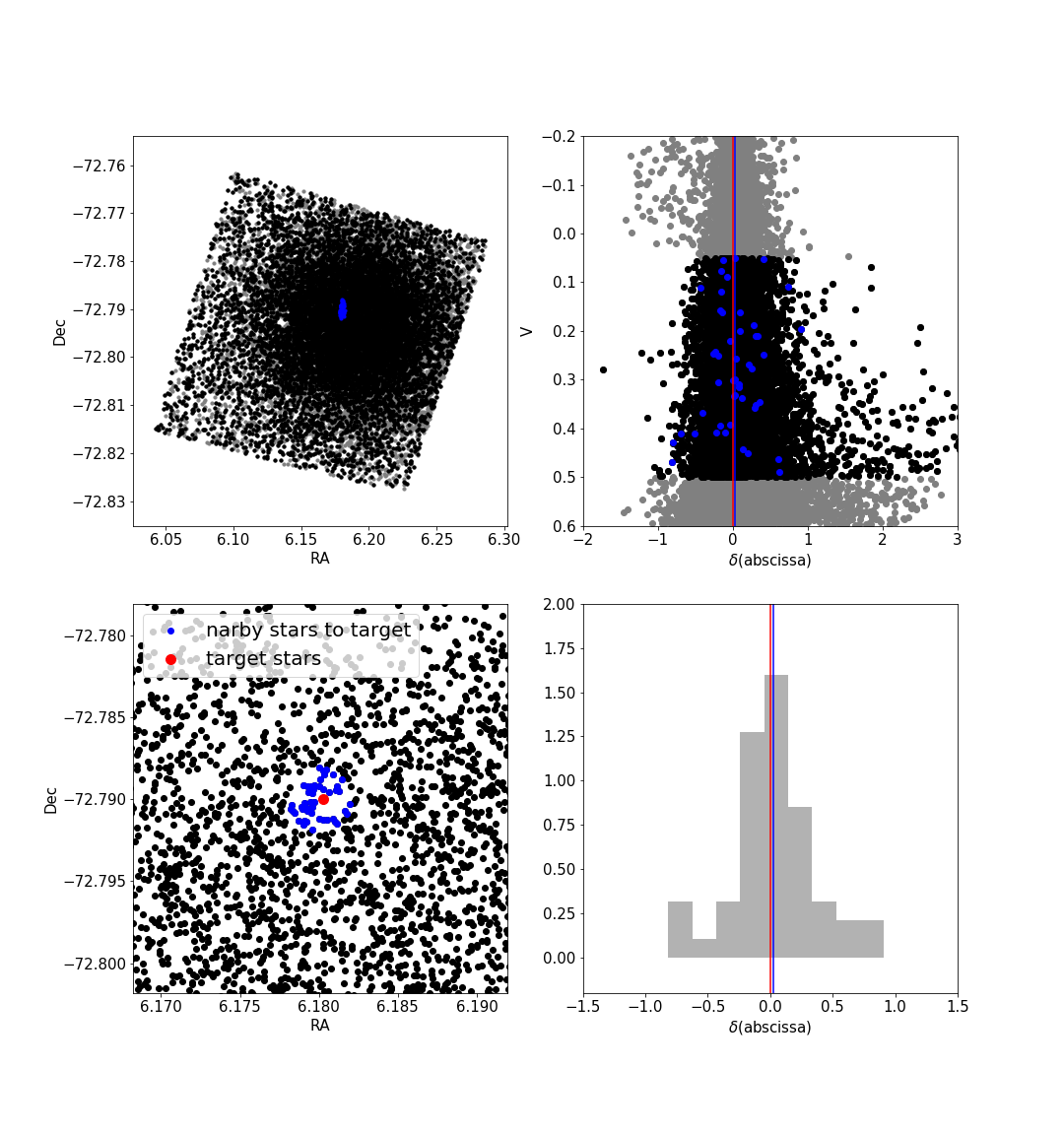}
\caption{(Top left) HST/ACS field of view of Kron 3 reference stars to calculate differential reddening. Blue points indicate nearby 49 stars around target star. (top right) ordinate vs $\delta$abscissa distribution of reference stars and 49 nearby stars. Red and blue lines are median $\delta$abscissa of reference stars and 49 nearby stars. 
(bottom left) zoomed version of top left plot. (bottom right) $\delta$abscissa histogram of 49 stars. Median $\delta$abscissa is also over plotted.}
\label{diff_red}
\end{figure*}


We have used only MS stars to generate red fiducial line. First, we binned the MS in 0.2 mag along ordinate and calculated median abscissa associated to median ordinate for each bins. The fiducial line is then created by fitting these points with a cubic spline. The use of median points help us to minimize the contamination effect from binary stars, field stars and stars with poor photometry. The red line in the middle panel of Figure \ref{cmd_rot} represents spline fit to the median points (blue) of each bin.
Then we calculated the distance to each star from the fiducial line along the reddening axis ($\delta$abscissa). The right panel of Figure \ref{cmd_rot} shows the plot of ordinate vs $\delta$abscissa. We choose these black points as references to calculate the differential reddening suffered by each stars in the ordinate vs $\delta$abscissa diagram.

The final step is explained in the Figure \ref{diff_red}. The idea is if the cluster does not effected by differential reddening then median value of stars from any small part of the cluster region in the $\delta(abscissa)$ vs ordinate diagram will coincide with the red fiducial line. On the other hand, if the cluster has differential reddening the median value will show an offset to the red fiducial line. Here we describe the procedure to calculate differential reddening suffered by a target star. 
We choose a small region around that target star in the spatial diagram and over plotted on the ordinate vs abscissa diagram. Then, we calculate the median value of $\delta$abscissa of these stars with respect to the fiducial line, which gives us the differential reddening associated to that star. The top left panel of Figure \ref{diff_red} shows the spatial distribution of reference stars from the ordinate vs $\delta$abscissa plot and the blue points indicates nearby 49 stars around the target star (red point) of interest, which is further zoomed in the bottom left panel. The top right panel shows  distribution of 49 stars (blue) in ordinate vs $\delta$abscissa diagram, over plotted on the distribution of all the reference stars. The blue line indicated the median value of $\delta$abscissa of that 49 stars which is slightly shifted from the median $\delta$abscissa (red line) of reference stars. The difference (0.025) provides us differential reddening suffered by the target star. The bottom right panel shows the histogram of $\delta$abscissa of that 49 stars. This step is repeated for all the reference stars to get the distribution of differential reddening suffered by all the reference stars across the field. 



\begin{figure}
\centering
\includegraphics[width=\columnwidth]{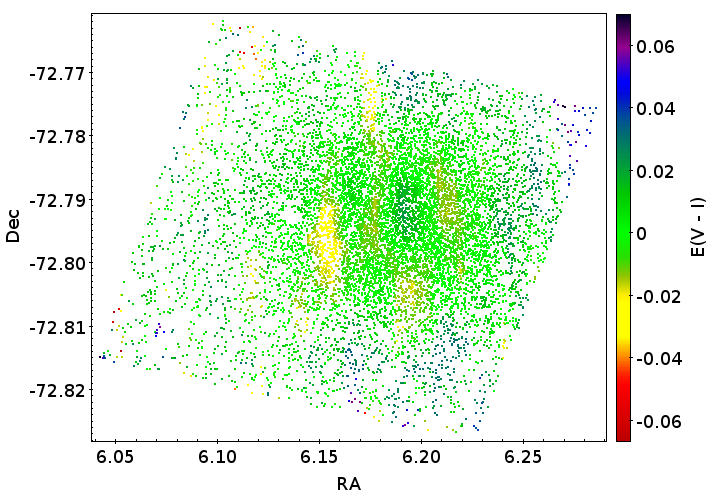}
\caption{Differential reddening map of Kron 3. }
\label{red_map}
\end{figure}

\begin{figure}
\centering
\includegraphics[width=\columnwidth]{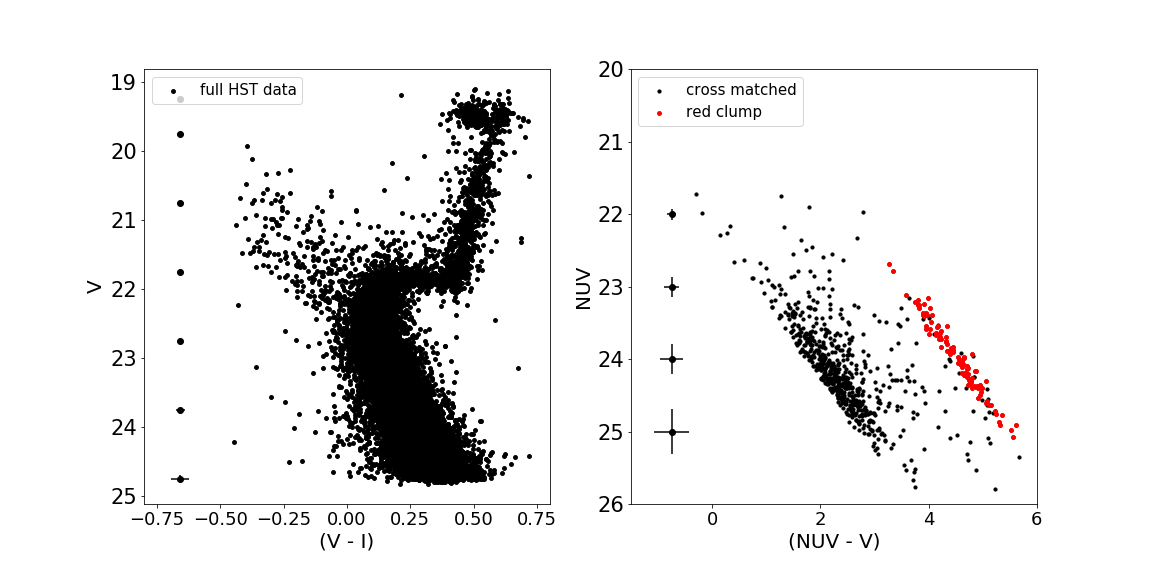}
\caption{Differential reddening corrected CMDs of Kron 3. }
\label{redd_corr_cmd}
\end{figure}

After we subtracted $\delta$abscissa to the abscissa of each star in the rotated CMD, we obtain an improved CMD. We have further used that CMD to to derive a more accurate selection of the sample of MS reference stars and derive a more precise fiducial line. We have iterated the procedure and after three iterations the procedure has converged. 
The improved abscissa and ordinate values are then converted to $V$ and $I$ magnitudes. Now the comparison between the original magnitudes with the corrected ones will provide us variation in $E(V-I)$ for each star. Figure \ref{red_map} shows the reddening map (variation in $\delta E(V-I)$) of the cluster. The maximum variation in differential reddening suffered by the stars is $\sim \pm$0.06. We have also shown differential reddening corrected CMDs in the Figure \ref{redd_corr_cmd}, which suggest that extension in RC stars are still present even after incorporating the effect of reddening variation.

\subsection{PSF variation}

As demonstrated in Sec. 4.1, Kron 3 exhibits very low levels of intrinsic and differential reddening. 
In this section, we check for spatial variation of the photometric zero point due to small, unmodelable PSF variations. As for the case of differential reddening, we adopted the method described by \citet{milone2012}, which suggests that the most evident manifestation is a shift in the color of the cluster sequence as a function of the location in the field (Anderson et al.2008). However, we have only one filter (NUV) with a sufficient number of sources to construct a UV CMD. So, we are unable to carry out the method for NUV CCD. Whereas, in recent calibration paper by \citet{tandon2020}, it is shown that for the SMC calibration field that there is no variation in PSF (FWHM of 2.63 sub-pixels or 1.08$''$) with 7.5 arcmin around the center of the UVIT CCD. Kron 3 is located within 3 arcmin of the center and as such we expect there to be no variation in the PSF.

The adopted method is applied to the HST data and is very similar to as described in section 4.1. The only difference is that CMD has not been rotated with respect to the reddening vector, so that we made the correction along the colour axis. All other steps are same. The maximum colour variation is around 0.01 mag for HST data, which corresponds to 0.03 mag in $NUV-V$ colour. So, the expected spread due to PSF variation is also significantly smaller than the observed spread seen in the RC stars in the UV.


We found that photometric zero point variations (differential reddening and/or PSF variation) cannot explain the extended RC in the UV-optical CMD of Kron 3. Now we have to examine whether field contamination could account for the observed spread. 
In order to ensure whether these bright stars are cluster members or not, it is important to remove field stars from the CMD. As the HST data covers only the central part of the cluster, it is not possible to remove field stars in optical or UV-optical CMDs. 
Hence, we used {\it Gaia} DR2 to isolate field stars in the cluster CMD, as described in the following section.


\section{UV-Optical CMD using UVIT and {\it Gaia} data}

We used the {\it Gaia} DR2 catalog and considered detections within 15$'$ radius around the cluster center to study the field star distribution and its effect on the cluster properties. 
{\it Gaia} DR2 sources have large astrometric errors in crowded fields such as in the central regions of dense clusters, which are measured by astrometric$\_$excess$\_$noise parameters \citep{lindegren2012,lindegren2018}. A zero value for this parameter indicates a reliable astrometric solution, while high values correlate with unreliable astrometric solutions. 
We selected stars with astrometric$\_$excess$\_$noise not exceeding a median value of 1.3 mas at G=19 mag; faint sources in crowded regions correspond to higher values.
A large fraction ($\sim$70\%) of stars is removed by this condition within the central $\sim$ 0.$'$5 radius of the cluster. Hence, {\it Gaia} DR2 is not useful to study the core of this cluster. 


\begin{figure}
\centering
\includegraphics[width=\columnwidth]{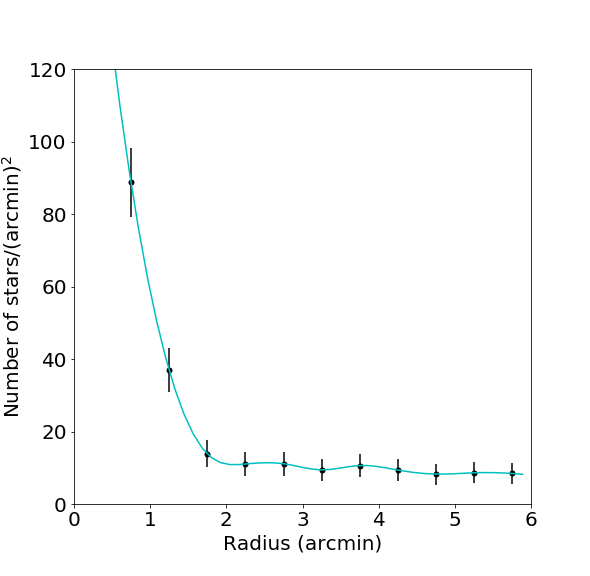}
\caption{Radial density profile of Kron 3 in steps of 0.$'$5 using Gaia DR2 data and fitted with a cubic spline function (cyan).}
\label{king_gaia}
\end{figure}

We have also used {\it Gaia} DR2 to determine the radius of the cluster. We calculated the radial density of stars (number of stars/area) with a bin width of 0.$'$5 and plotted it against radius in Figure \ref{king_gaia}. We fitted the data points (black dots) with cubic spline (cyan line) and found that the density of stars decreased with radius and became equal to the  density of field stars at $\sim$2.$'$0.   
Therefore, the radius of the cluster based on optical Gaia DR2 is similar to the value obtained from UVIT-NUV data.


\begin{figure}
\centering
\includegraphics[width=\columnwidth]{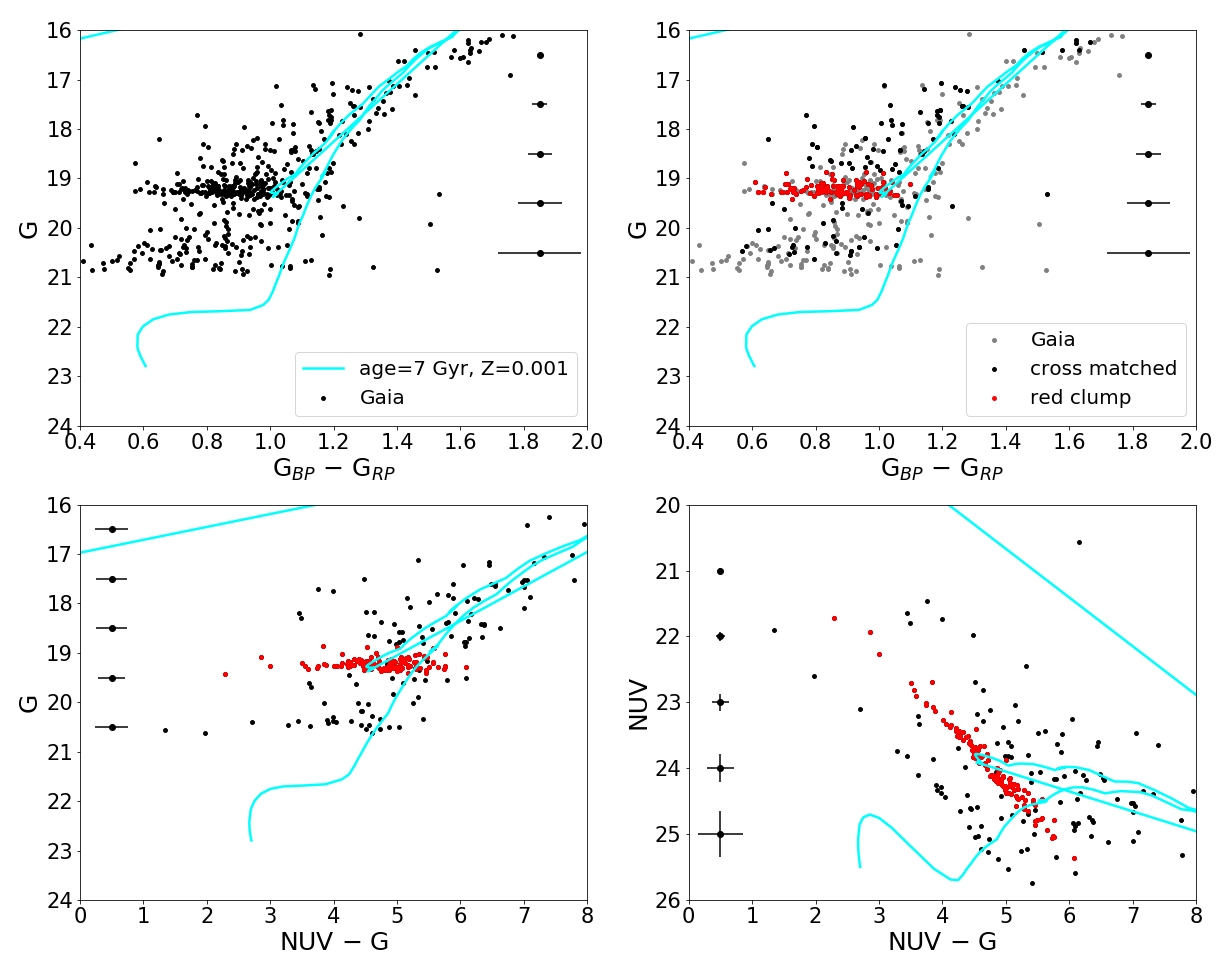}
\caption{(top-left) CMD of Gaia DR2 stars within the central 2’ of Kron 3. (top-right) CMD of Gaia DR2 stars cross-matched with NUV data (black) over-plotted onto the full Gaia DR2 sample (grey) with red clump stars indicated in red. The same stars are plotted in the G vs (NUV-G) CMD (bottom-left) and in the NUV vs (NUV-G) CMD (bottom-right). The isochrone plotted in all panels is as in Figure \ref{cmd_hst}.}
\label{cmd_gaia}
\end{figure}


In the top-left panel of Figure \ref{cmd_gaia} we plotted G vs ($G_{\rm{BP}}-G_{\rm{RP}}$) CMD for the cluster region (stars within a radius of 2.$'$0 around the center). An isochrone of age 7 Gyr and metallicity Z=0.001 was also plotted onto the CMD after  correcting for DM and reddening. 
In the right-panel of Figure \ref{cmd_gaia}, we over-plotted G vs ($G_{\rm{BP}}-G_{\rm{RP}}$) CMD for cross-matched {\it Gaia} stars (black points) on top of the CMD of {\it Gaia} stars (grey points) in the cluster region; RC stars are marked in red. 
The isochrone was also plotted onto the CMD as in the previous panel. RC stars extend over 0.4 mag in the colour axis in both figures. 
The bottom-left panel of Figure \ref{cmd_gaia} shows the G vs (NUV$-$G) CMD of the cross-matched Gaia-UVIT stars (238) where RC stars spread over two magnitudes in colour. The criteria used to cross-match the Gaia-UVIT data is similar to that obtain HLA-UVIT cross-matched data. 
The bottom-right panel shows the NUV vs (NUV$-$G) CMD for the same stars. Here, RC stars exhibit a similar spread in both colour and magnitude. 
Thus, a single isochrone is not able to fit the colour extension of RC stars. To investigate whether the extension is due to field star contamination, we statistically decontaminate the field stars as follows.


\begin{figure}
\centering
\includegraphics[width=\columnwidth]{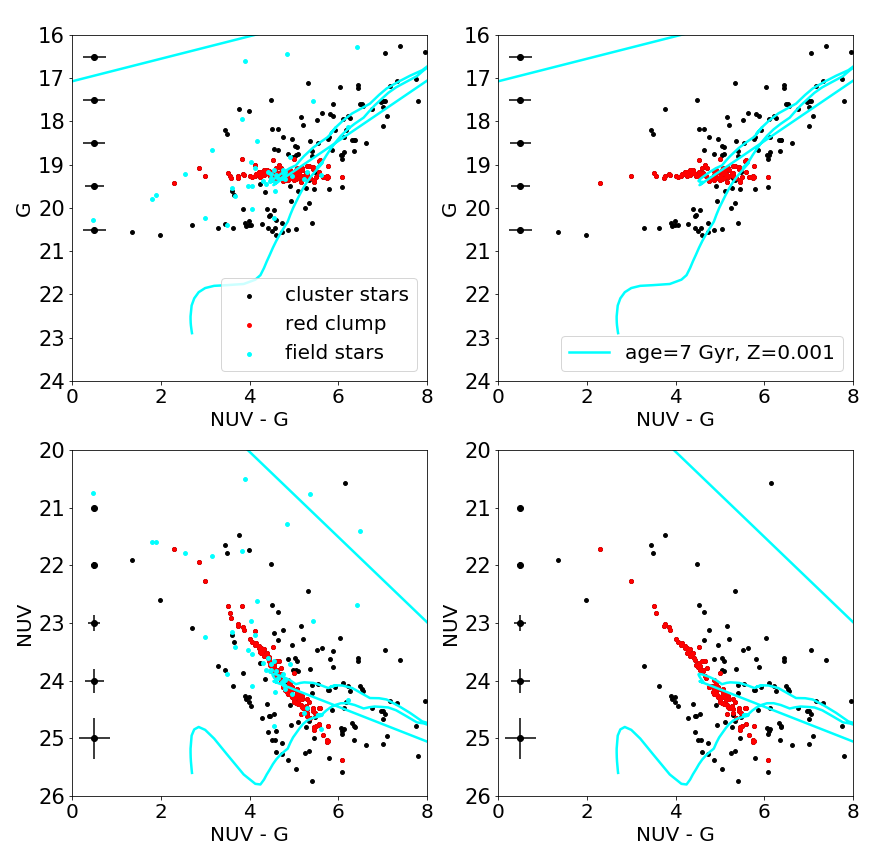}
\caption{CMDs of stars within the central 2’ of Kron 3 cross-matched between Gaia DR2 and NUV data for the full sample (left) and for only cluster stars where field stars have been removed (right). CMDs for both G vs ($G_{\rm{BP}}-G_{\rm{RP}}$) and NUV vs ($G_{\rm{BP}}-G_{\rm{RP}}$) are shown in top and bottom rows, respectively. The isochrone plotted in all panels is as in Fig. \ref{cmd_hst}.}
\label{field_sub}
\end{figure}


\subsection{Removal of field star from cluster CMDs} 

To carry out a statistical process, we considered that field stars are uniformly distributed in the sky over a few arcmin around the cluster. An annular field region, with an inner radius of 3$'$.0 and an area equal to that of the cluster region was chosen outside the cluster to carry out the process. 
We first constructed G vs (NUV$-$G) CMD for the cluster and field regions. 
The field stars within the cluster region are then removed by considering each star in the field CMD and finding its nearest counterpart in the cluster CMD. We considered a grid of [magnitude, colour] bins with different sizes, starting with [$\Delta$G , $\Delta$(NUV $-$ G)] = [0.02, 0.01] up to a maximum of [0.5, 0.25], where the units are in magnitude. 

In the top-left panel of Figure \ref{field_sub}, we plotted the G vs (NUV$-$G) CMD for the cluster (black) and field region (cyan); RC stars are marked in red. The top-right panel shows the CMD of the cluster after subtracting field stars. 
We notice the presence of an extended RC distribution even after the removal of field stars. 
The bottom panels show NUV vs (NUV-G) CMDs for the same stars as in the top panels. This figure suggests that the extension of the RC is not due to field contamination, rather it is an intrinsic property of the cluster. It is therefore essential to check all the intrinsic properties which can contribute to make the RC stars brighter and hotter in UV-optical CMD.

\section{Possible reasons behind the extended RC}

As the RC stars are found to be members of the cluster and the observed spread in RC is an intrinsic property of the cluster, then the possible reasons for the extended RC could be (i) variable mass loss, (ii) presence of age and metallicity gradient among the stars, (iii) presence of multiple stellar population due to variation in the light elemental abundances (CNO), (iv) variation in the initial helium abundances within the cluster. 

Stars with higher mass loss rate can expose their inner hotter layer by expelling the outer layer and get brighter in NUV. Stars with different metallicity values will appear in different location in the CMD. The metal poor stars will appear brighter and bluer in the CMD whereas the metal rich stars will appear relatively fainter and redder. So, the presence of metallicity gradient can cause the extended feature in the RC. This feature becomes more prominent in the UV-optical CMD as the UV region is more sensitive to metallicity. Study by \citealp{holly2018} showed the presence of variation in C and N abundances in the RGB stars of this clusters. RC stars are not yet been reported for the presence of CNO variation. Presence of molecular bands (CN, NH, CH) in the NUV region can show the variation in NUV flux due to variation in the C-N-O abundances, which may produce the observed spread. Massive clusters ($> 10^5 M_\odot$) in the Magellanic Clouds are found to host 2G stars with enrichment in initial helium abundances \citep{chantereau2019}. Here, we did not able to separate two distinct population but rather observed a spread in the CMD. It will be interesting to check if helium can be one of the reason behind the observed spread for this massive SMC cluster.  
Hence, there are more than one possibilities to get a large spread in the RC.
We have analyzed all the above mentioned possibilities to ensure we understand the reason for getting an extended RC in the cluster Kron 3.

\subsection{Different mass loss}

\begin{figure}
\centering
\includegraphics[width=\columnwidth]{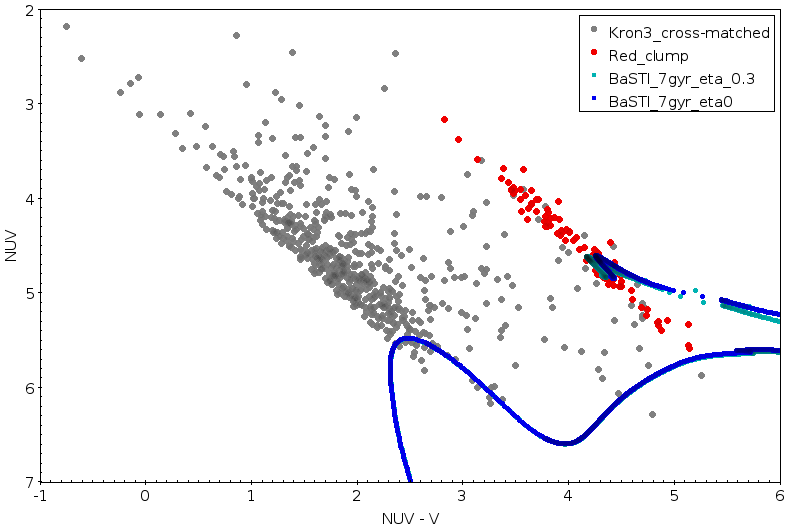}
\caption{NUU vs (NUV-V) CMD for HLA-UVIT cross-match data. BaSTI isochrones for different mass loss rate with fixed age has been over plotted on CMD.}
\label{mass_loss}
\end{figure}

\citealp{tailo2019, tailo2020} showed that the mass loss during the red giant branch(RGB) phase is an fundamental ingredients to constrain the horizontal-branch (HB) morphology in the globular clusters (GCs). 
Therefore, we have also taken into account the effect of mass loss in the our study to check if mass loss can be one of the cause in the extended features of RC. We have generated synthetic CMDs for two different mass loss rate using BaSTI model for F555W and N242W filters and over plotted on observed NUV-optical CMD after correcting for reddening and distance modulus. In Figure \ref{mass_loss}, blue and Cyan points shows two synthetic CMDs for mass loss rate 0 and 0.3, respectively. We notice that change in the location of RC point is very small compare to observed spread. Therefore, we suggest that different mass loss rate may not be reason behind the observed extension in NUV – V colour.

\begin{figure*}
\centering
\includegraphics[width=1.8\columnwidth]{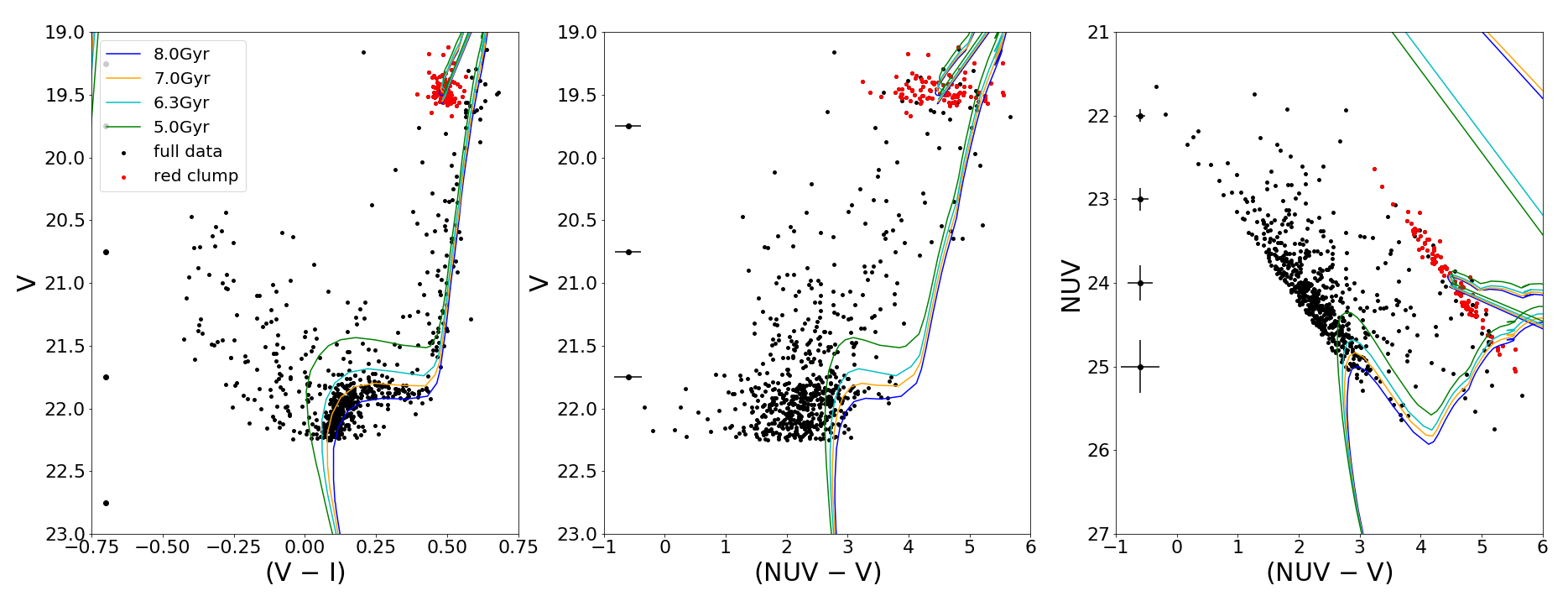}
\caption{CMDs of the HLA-UVIT cross-matched sample within 2.$'$0 of Kron3. From left to right they are V vs ($V-I$), V vs ($NUV-V$) and NUV vs ($NUV-V$), respectively. Red points denote the red clump stars. Isochrones for a range of ages (5-8 Gyr, marked with different colours) are over-plotted onto each CMD with the metallicity fixed at Z=0.001.}
\label{diff_age}
\end{figure*}

\begin{figure*}
\centering
\includegraphics[width=1.8\columnwidth]{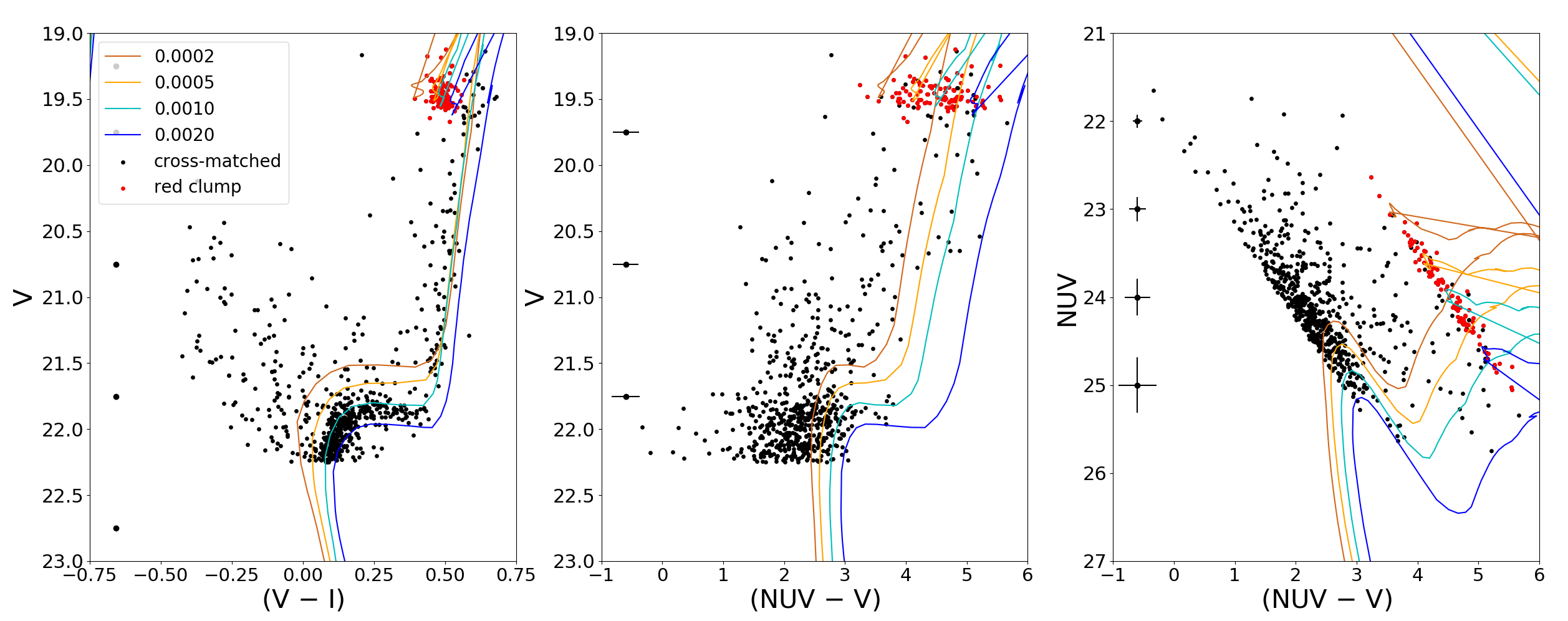}
\caption{The same as Fig. \ref{diff_age} but over plotting isochrones with a range of metallicities (Z=0.0002-0.002, marked with different colours) at a constant age of 7 Gyr.}
\label{diff_met}
\end{figure*}


\subsection{Age and metallicity spread in Kron 3}
\subsubsection{colour of RC stars based on stellar evolutionary models}

To check whether the age and metallicity variation in Kron 3 can explain the spread in the RC in the NUV, we used HLA-UVIT cross-matched data. In Figure \ref{diff_age} we plot V vs ($V-I$), V vs ($NUV-V$) and NUV vs ($NUV-V$) CMDs highlighting RC stars. Then, we over-plotted isochrones on each CMD with ages ranging from 5 to 8 Gyr, keeping the metallicity value constant at $Z$=0.001. 
This figure suggests that although the age spread can explain the broad MS in the V vs ($V-I$) CMD, it is unable to fit the MS spread in both the V vs ($NUV-V$) and NUV vs ($NUV-V$) CMDs. The (V vs $V-I$) CMD suggests that the cluster has a spread in age from 6 to 8 Gyr. 
Despite the difference in age, the positions of the RC from the various isochrones are almost identical in both colour and magnitude, and thus a simple age spread is unable to account for the observed spread in the RC stars in the NUV.

In Figure \ref{diff_met}, we plot the same CMDs as shown in Fig. \ref{diff_age}, however we now show isochrones of the same age, but with different metallicities ranging from Z=0.0002 to 0.002. 
The Figure shows that isochrones of different metallicities are able to reproduce the observed spread in the RC. We suggest that a spread in metallicity from Z=0.0002 to 0.002 is likely to be one of the reason behind the extended RC in the NUV.

A similar spread in the NUV$-$optical colour has also been observed among RC stars in the Milky Way, which  appears to be strongly correlated with spectroscopic metallicities \citep{mohammed2019}. These authors used a sample of 5175 RC stars, observed as part of the Sloan Digital Sky Survey Apache Point Observatory Galactic Evolution Experiment (APOGEE) and combined these with GALEX All Sky Imaging Survey (GIAS; \citealp{martin2005}) and Gaia DR2 photometry, to construct NUV-optical CMDs of RC stars. 
They aimed to establish a relation between absolute (NUV$-$optical) colour and the spectroscopic metallicity in order to photometrically determine the metallicity of any RC star, whose distance is known. They noticed that a spread of 0.7 mag in absolute ($G_{\rm{BP}}-G_{\rm{RP}}$) increased to over 4 mag in absolute NUV$-$G colour. They found a strong correlation between (NUV$-$G) and [Fe/H] with a standard deviation of about 0.16 mag. Using stellar evolutionary models from MESA Isochrones and Stellar Tracks (MIST) project, \citet{mohammed2019} also showed that the relation between NUV$-$G colour and metallicity for RC stars closely matches the observed trend and it does not depend much on the initial mass and age. Hence, this relation can be used to estimate photometric metallicities where a spectroscopic metallicity measurement is missing. 
This study supports the suggestion that the extension of RC stars of Kron 3 in the UV-optical CMD, is probably due to the presence of a metallicity range within the cluster.  
%


\begin{figure*}
\centering
\includegraphics[width=\columnwidth]{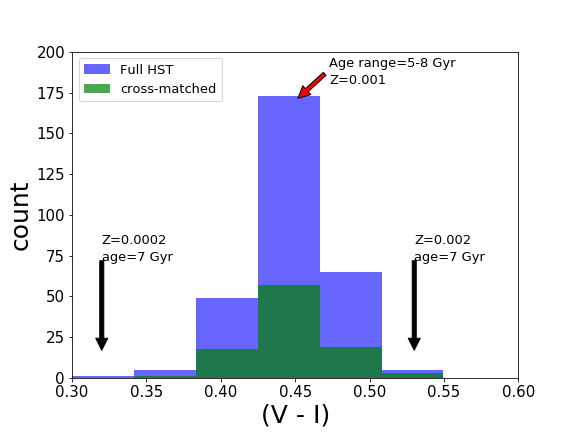}    \includegraphics[width=\columnwidth]{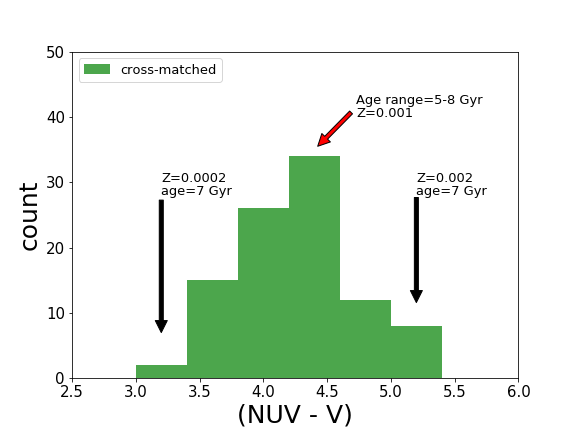}
\caption{Distributions of RC stars as a function of ($V-I$) colour (left) and ($NUV-V$) colour (right) for the HLA data (blue) and the HLA-UVIT cross-matched data (green).
}
\label{hist}
\end{figure*}


In Figure \ref{hist}, we plot the distribution of RC stars against colour. The left-panel shows the histogram of ($V-I$) colour of RC stars with a bin size of 0.04 mag for full HLA data of Kron 3 and HLA-UVIT cross-matched data, denoted in blue and green colour respectively. Both distributions peak at $V-I$ $\sim$0.44 mag. The peak of the distribution corresponds to isochrones within an age range of 5-8 Gyr and Z=0.001 (red arrow) while bins corresponding to isochrones with an age of 7 Gyr and a metallicity Z=0.0002 and 0.002 are indicated with black arrows. 
The right-panel shows the histogram of (NUV-V) colour of RC stars with a bin size of 0.4 mag for the HLA-UVIT cross-matched sample. The distribution peaks at (NUV-$V$) $\sim 4.4$\,mag and arrows indicate specific bins as above. 


\begin{figure}
\centering
\includegraphics[width=\columnwidth]{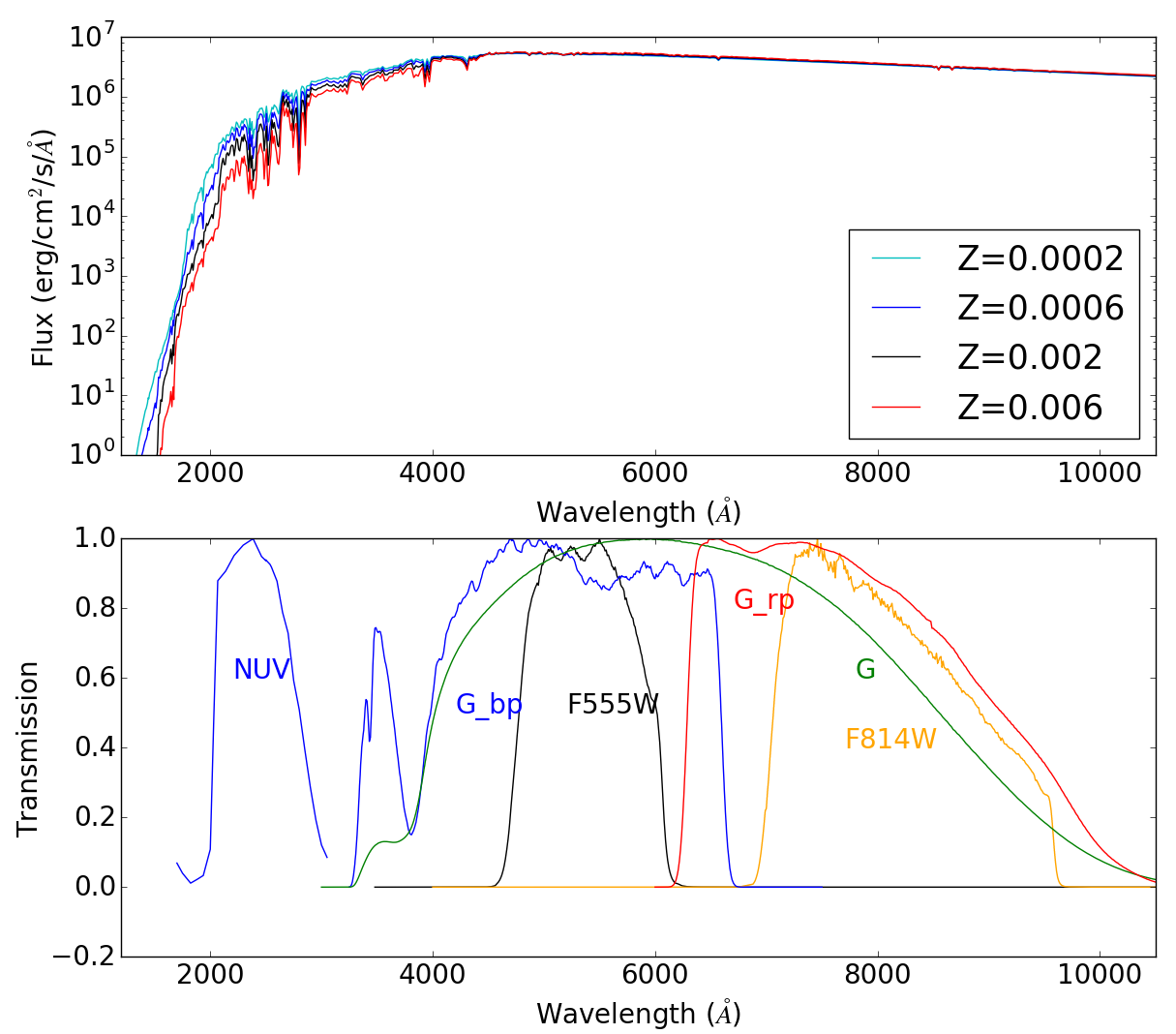}
\caption{Model spectra for a metallicity from Z=0.0002 to 0.006 (marked in different colours) are presented in the upper panel, for T$_{eff}$=5250 K and log(g)=2.5. In the lower panel the plot shows response curves for pass bands used in this study as indicated. }
\label{flux_filter}
\end{figure}


\subsubsection{Colour of RC stars based on model spectra}

\begin{figure}
\centering
\includegraphics[width=\columnwidth]{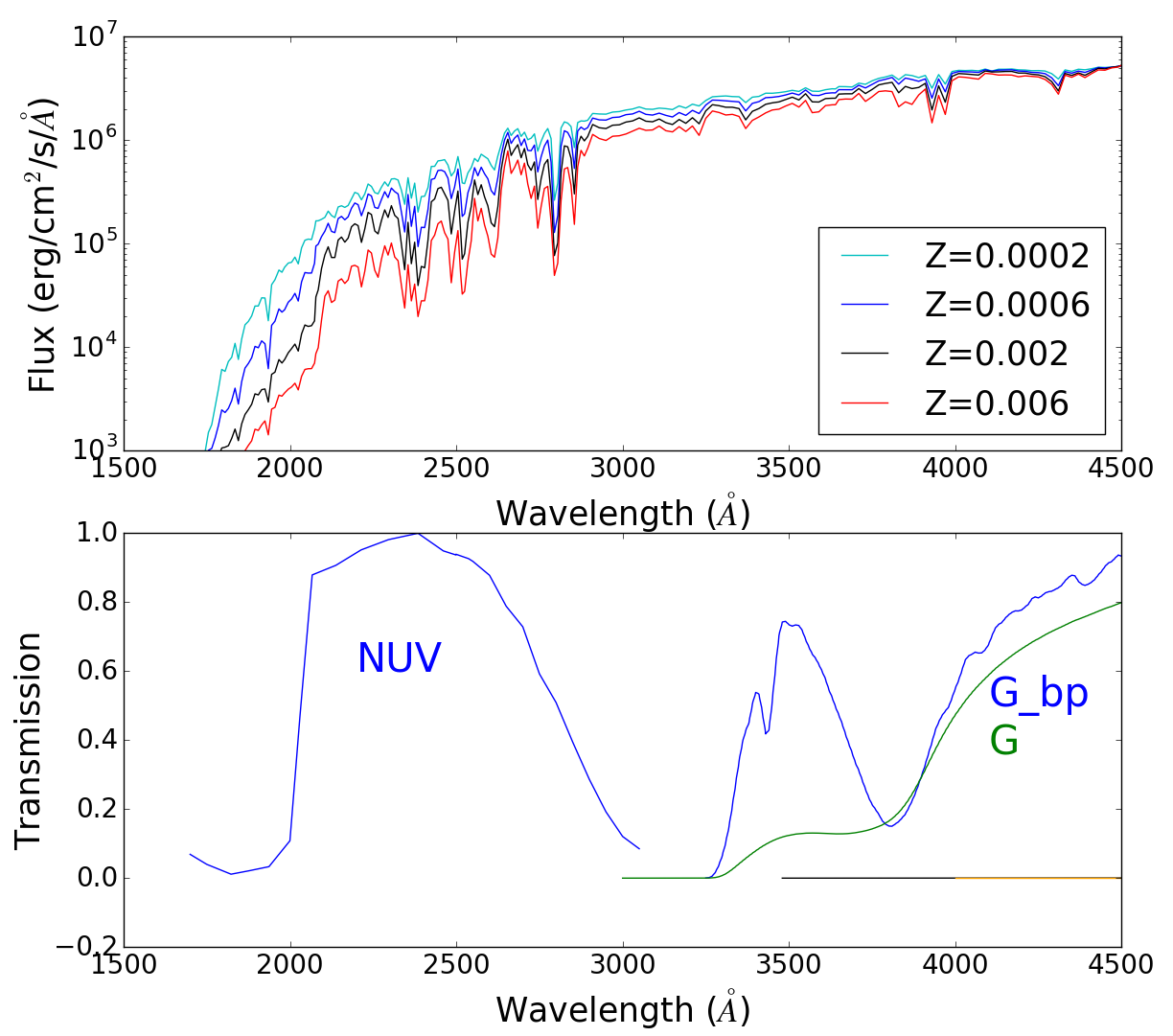}
\caption{Zoomed-in version of Fig. \ref{flux_filter} to show the absorption lines and flux variation prominently below 400 nm for different metallicities.}
\label{flux_zoom}
\end{figure}

If there is a metallicity gradient among the RC stars, then the NUV flux of metal poor stars should be higher than that of the metal rich stars. 
It is therefore necessary to calculate the expected flux for RC stars in different bands from theoretical synthetic spectra and to determine whether there is excess flux in the NUV within the metallicity range mentioned above. We used the \textsc{atlas9} model spectra \citep{cas1997} obtained from the Virtual Observatory Spectral Energy Distribution (SED) Analyzer  (VOSA; \citealp{bayo2008}) tool to estimate the flux in different bands.  
The model spectra cover a large range of the following parameters: the metallicity ranges from [Fe/H]=$-$2.5 to 0.5 dex, log(g) ranges from 0 to 5.0 and T$_{eff}$ ranges from 3500 to 50000 K. [Fe/H] and log(g) vary in steps of 0.5 whereas T$_{eff}$  varies in steps of 250 K. In the isochrone table we found that RC stars have a surface gravity value log(g) $\sim$ 2.5 which does not change much with respect to age, metallicity or temperature. We also found that RC stars have a temperature range T$_{eff}$ = 5100-5600 K for a  metallicity range Z=0.0002-0.002 at an age of 7 Gyr. 
In Figure 14 we plot model spectra covering a range of metallicities, assuming a fixed value for the effective temperature ($T_{\rm{eff}}=5250$\,K) and surface gravity (log$g$=2.5\,dex). Figure \ref{flux_filter} shows that there is a significant difference in flux for different metallicities at wavelengths below 400 nm, which is not prominent at longer wavelengths. 
In the lower panel we showed normalised response curves for different filters used in this study, taken from Spanish Virtual Observatory (SVO) filter profile service. 
Note that the NUV filter N242W falls within the region where the variation in flux is noticeable. The zoomed-in version of this region is shown in Figure \ref{flux_zoom}, to highlight the flux variation in the above mentioned metallicity range. It is apparent that the continuum flux reduces to almost half when the metallicity changes from Z=0.0002 to 0.002 at 5250 K. A flux decrease of a factor of two, corresponds to a magnitude difference of 0.75\,mag. We also noticed that absorption spectral lines get deeper with increasing metallicity causing a reduction in flux. Figures \ref{flux_filter} and \ref{flux_zoom} suggest that the UV region is more sensitive to metallicity than optical and near-IR regions. As most of the absorption lines due to metals appear in the UV, the higher the metallicity the greater the absorption by metals present at the surface of stars which will make it fainter. As there are more metallic lines present in the UV region than in the optical or near-IR regions, the variation in flux due to metallicity is more prominent in the UV region.  
This could be one of the reasons for some of the RC stars getting brighter in the NUV due to a low metallicity, which contributes to extending the RC in the NUV-optical CMDs, but not in optical CMDs.


\begin{figure*}
\centering
\includegraphics[width=1.8\columnwidth]{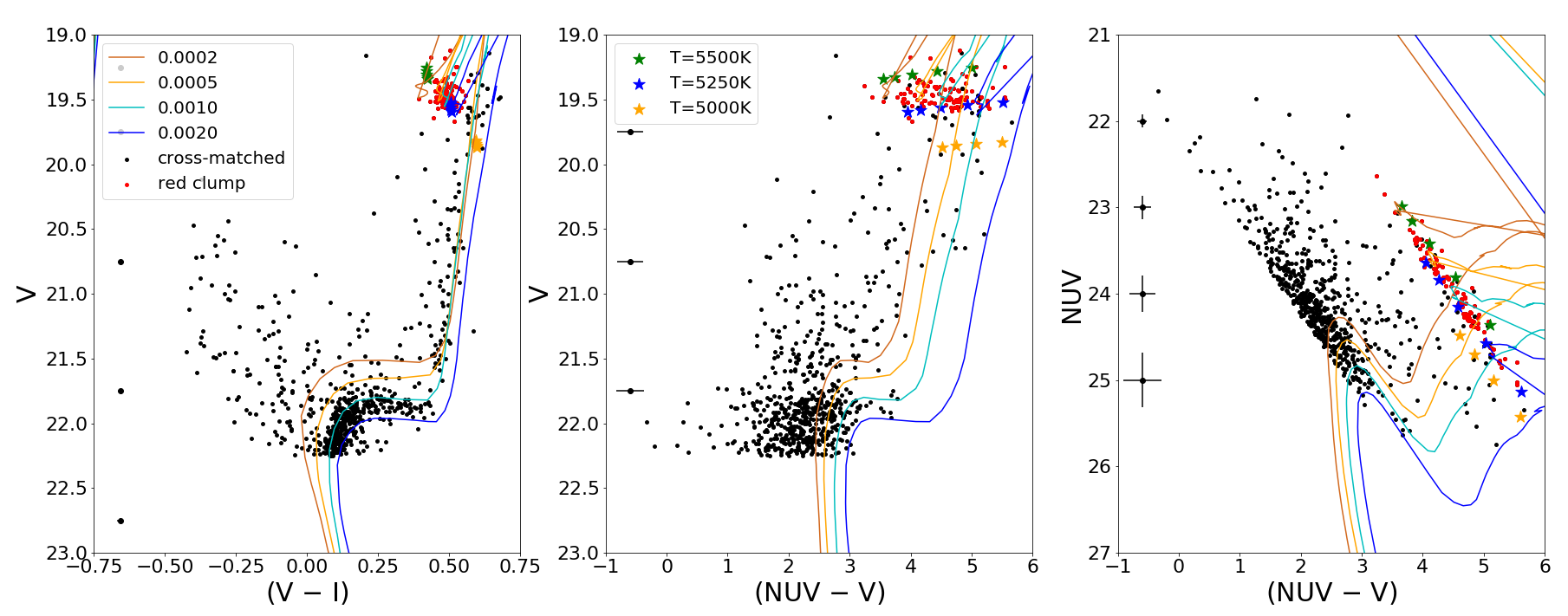}
\caption{The same as Fig. \ref{diff_met} but for a range of metallicity Z=0.0002-0.002 over plotting also the expected colours and magnitudes for RC stars for different temperatures (5000, 5250 and 5500 denoted by orange, blue and green points, respectively) and metallicities (Z=0.0002, 0.0006, 0.002 and 0.006) after correcting for DM and reddening.}
\label{model_rc}
\end{figure*}


To get a quantitative estimate of the expected flux in all pass bands, we convolved the filter responses with the model spectra for different temperatures and metallicities using the VOSA tool. We chose a temperature range 5000-5500 K and a [Fe/H] from $-$2.5 to $-$0.5 dex, to estimate fluxes in different pass bands. Then, we scaled the flux to a distance of 10 pc by multiplying it with a scaling factor (R$_c$/D)$^2$ to calculate the absolute flux, where R$_c$ is the average radius and D the distance (10 pc) of RC stars. The absolute flux was converted to absolute magnitude ($M$) using the standard relation, $M$ = $-$2.5 log(F$_*$/F$_0$), where  F$_*$ is the flux of the star and F$_0$ is the zero-point flux in a given filter. Zero-point flux values were are provided as part of VOSA. In Figure \ref{model_rc} we overlay the model-generated colours and magnitudes of RC stars after accounting for DM and reddening. The green points indicate the magnitude obtained for different metallicities at a temperature of 5500 K. 
The blue and orange points denote the same for temperatures of 5250 and 5000 K, respectively. We also overlay isochrones for a range of metallicity (the same values as discussed earlier when describing Fig. \ref{diff_met}) but at a fixed age of 7 Gyr. The observed RC clump stars are marked in red. The left panel shows the optical CMD (V vs $V-I$) using HLA data. The figure also shows that model generated magnitudes are clumped together for a constant temperature, suggesting that the RC population is not stretched in either colour or magnitude, due to metallicity variations. On the other hand, model generated magnitudes appear stretched in ($NUV-V$) colour at a constant temperature and show only a small variation in magnitude. 
The middle panel shows that the range of temperatures and metallicities can describe the extension of the RC, which gets more prominent in the NUV vs ($NUV-V$) CMD (right panel). In this panel, the extensions of the RC stars in both colour and magnitude are represented well by the theoretical models. 
This analysis supports and strengthens the suggestion that RC stars in Kron 3 have a range of metallicities.

\subsubsection{Spectral energy distributions of RC stars}

To verify that Kron 3 contains RC stars with a range of metallicities, we fit the optical/near-IR spectral energy distributions (SEDs) of a sample of Kron 3 RC stars using VOSA. 
This tool performs multiple iterations to fit the observed flux distribution with the theoretical model flux for different combinations of T$_{eff}$, log(g), [Fe/H] and $M_d$ values, and gives the best fitted parameters after performing a ${\chi}^2$ minimization. The scaling factor ($M_d$) is used to scale the model flux to match the observed flux and is defined as $(R_c/D)^2$ where $R_c$ is the radius of the star and D is its distance. 
We provided the extinction value of the cluster, A$_V$ = 0.082 \citep{glatt2008} as an input parameter to the VOSA tool. In this study we determined reduced ${\chi}^2$, which is defined as 
\begin{equation}
 {\chi}_{reduced}^2=\frac{1}{N-n}\sum_{k=1}^{N} \frac{(F_{o,k} - M_d \times F_{m,k})^2}{{\sigma_{o,k}}^2}\
\end{equation}
where N is the number of photometric data points, n is the number of input free parameters, $F_{o,k}$ is the observed flux and $F_{m,k}$ is the model flux. 
To obtain a better estimate of parameters, it is necessary to have many data points covering a wide range of wavelengths to generate SEDs and compare them with theoretical models. 
Therefore, we included near-IR data from the VISTA survey of the Magellanic Clouds system (VMC; \citealp{cioni2011}).

\begin{figure*}
\centering
\includegraphics[width=\columnwidth]{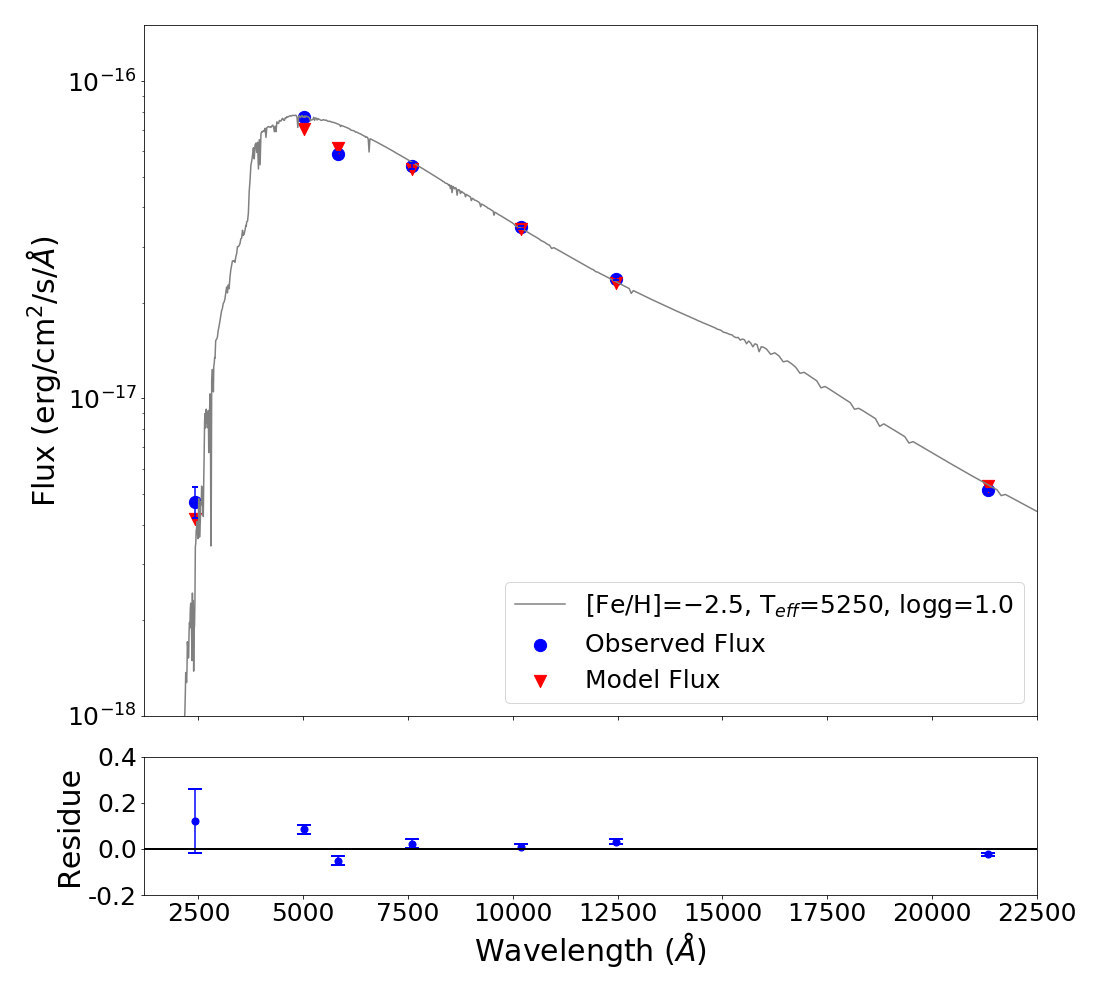} \includegraphics[width=\columnwidth]{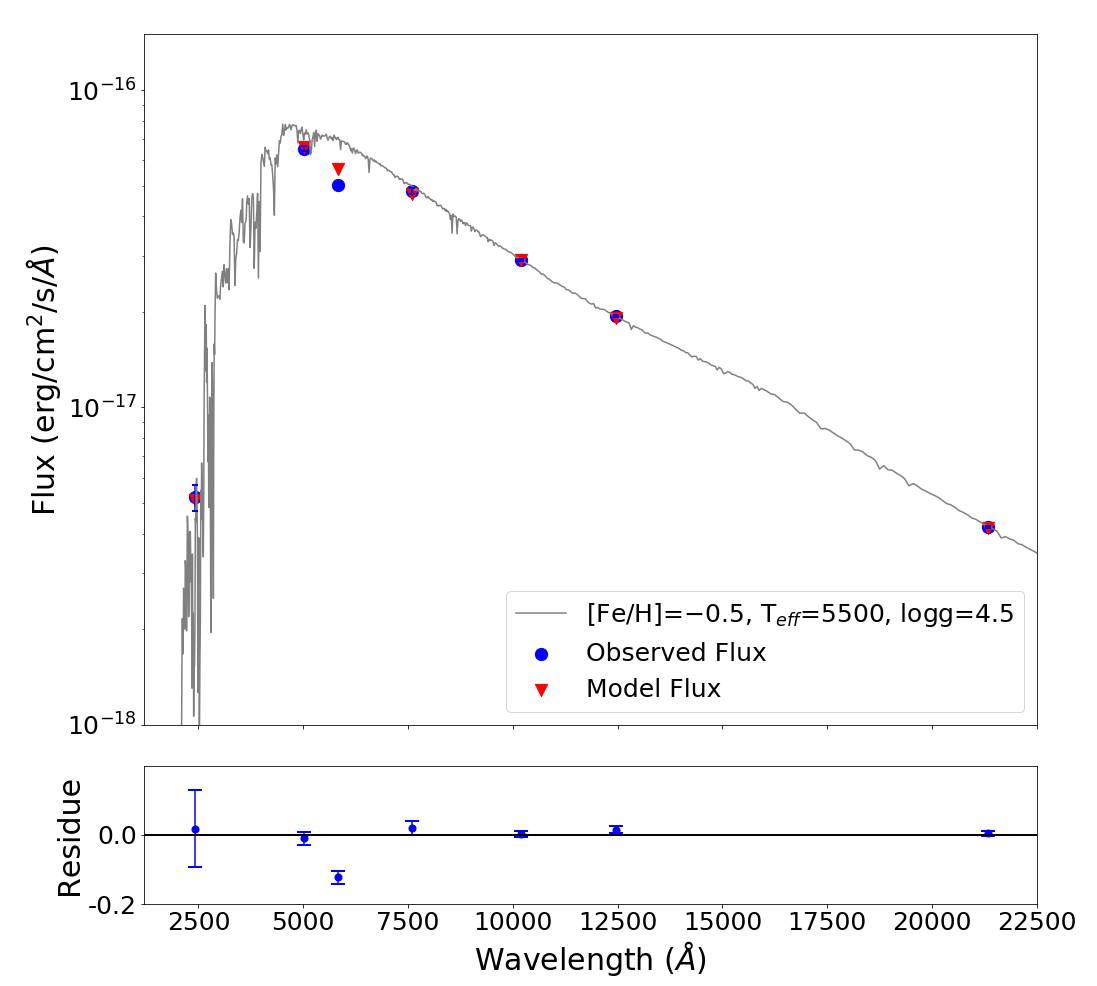}
\caption{SEDs of RC stars are presented in UVIT, Gaia DR2 and VMC data.
The blue and red points represent the observed and expected fluxes, respectively in different passbands. }
\label{sed1}
\end{figure*}

We used RC stars detected in all wavebands from UVIT (NUV), HLA, Gaia DR2 and VMC. We found 18 such stars and generated their SEDs after converting magnitudes to fluxes. We fitted these SEDs with the same \textsc{atlas9} model spectra as discussed in Sec. 6.2.2. 
The {\it Gaia}, UVIT and VMC fluxes were fitted well with the spectra.  
As the HST wavelengths are covered by the {\it Gaia} pass bands, we excluded the HST fluxes in the fits. Initially we set all the input parameters ([Fe/H], log(g) and temperature) as free values to obtain the best fit parameters. 
All the stars were fitted with a single spectrum. We found that for most of the stars the parameters lie between $-$2.5 to $-$0.5 dex for [Fe/H] and between 0 to 5 for log(g). For two stars [Fe/H] was found to be solar. Temperatures of all the stars were found to lie between 5250 and 5500 K. We have also found the values of luminosity and radius of these stars from the fitted spectrum, which lie in the range $\sim$60-90 L$_{\odot}$ and $\sim$8-11 R$_{\odot}$, respectively. Estimated values of effective temperature, luminosity and radius from SEDs are consistent with those expected of RC stars \citep{gallenne2018, wan2015}. 
In Figure \ref{sed1} we show model SED fits for two of these stars. The observed and expected model fluxes are denoted in blue and red, respectively, on top of the best fit theoretical spectrum (grey). 
In the left panel the observed fluxes of the RC star in different bands are fitted with a model spectrum of [Fe/H], log(g) and T$_{eff}$ values of $-$2.5, 1.0 and 5250K, respectively. In the right panel, the fitted model spectrum gives the values of [Fe/H], log(g) and T$_{eff}$ as $-$0.5, 4.5 and 5500K, respectively. The figure suggests a good match between the expected and observed fluxes within the uncertainties. 
In the bottom panel of figure \ref{sed1} we have also plotted the residuals ((observed flux $-$ model flux)/observed flux) for all filters, which appear all close to zero. 
${\chi}_{reduced}^2$ values for the fitted SEDs presented in Figure \ref{sed1} are found to be 18.1 and 10.6. 
The model fitted SEDs of the rest of the 16 stars are presented in the appendix in Figures \ref{sed3} and \ref{sed4}. The values of model fitted parameters are noted in those figures.


\begin{figure}
\centering
\includegraphics[width=\columnwidth]{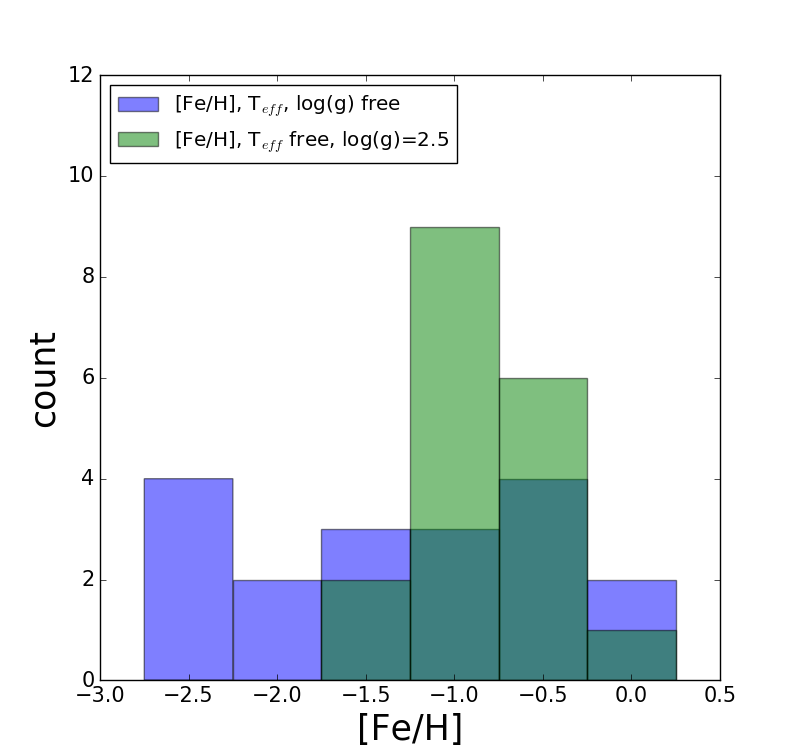}
\caption{Distribution of metallicity ([Fe/H]) values for RC stars obtained from the fit of their SEDs with theoretical models when metallicity, temperature and gravity are allowed to vary (blue) and when metallicity and temperature vary while gravity is fixed at log(g)=2.5 (green). 
}
\label{met_hist}
\end{figure}

Theoretical stellar evolutionary models suggest that RC stars have a surface gravity of ~2.5 dex and that this values is insensitive to changes in both age and metallicity. 
Therefore, we tried to fit the SEDs keeping log(g) fixed at 2.5 and leaving [Fe/H] and T$_{eff}$ as free parameters. We obtained that the best fit spectra for RC stars span a smaller range of metallicities, from $-$1.5 to 0 dex, while the temperature range is similar to the previous case where all three parameters were left to vary. 
We noticed that the ${\chi}_{reduced}^2$ values did not show much variation with respect to the previous estimates except for two stars. 
In Figure \ref{met_hist}, we show the distribution of [Fe/H] derived using SED fitting for 18 stars. The blue histogram represents the distribution of [Fe/H] when all the input parameters are left free to vary, whereas the green histogram shows the distribution of [Fe/H] when log(g)=2.5 and both [Fe/H] and T$_{eff}$ are left to vary. The distribution of [Fe/H] spans a range of values from $-$2.5 to 0 dex if we do not fix log(g), while fixing the value of log(g) gives a smaller range of [Fe/H], from $-$1.5 to 0 dex with a peak at $-$1.0 dex.

%

\subsection{Elemental abundance variation}

Almost all GCs shows split or broad red HB, that is mainly due to the combined effects of variations in light elemental abundances and helium. CN, CH and NH bands coincide with different photometric filters, and stars with different level of C and N exhibit flux variations in those filters, thereby leading to distinct populations split in colour and magnitude \citep{milone2020_wcen,milone2012_47tuc, piotto2015}. 
\citet{lagioia2019} found helium Variation in four Small Magellanic Cloud Globular Clusters. \citet{chantereau2019} found a spread in helium abundance in an intermediate age SMC cluster, Lindsay 1. 
Given that we have demonstrated the extended nature of RC stars in Kron 3, when viewed in the UV-opitcal CMD, we need to investigate whether variations in both the elemental and initial helium abundances could produce such a spread.


\begin{figure}
\centering
\includegraphics[width=\columnwidth]{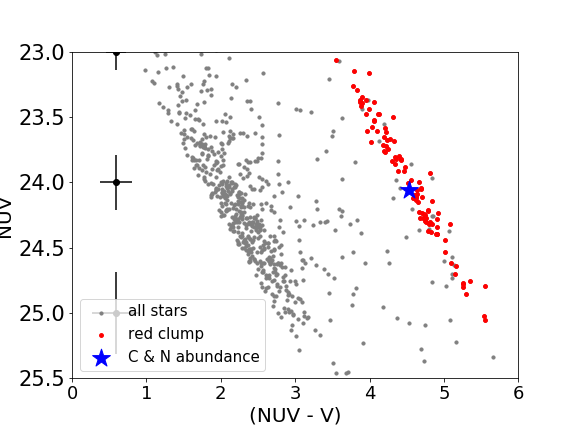}
\caption{NUV vs ($NUV-V$) CMD of the HLA-UVIT cross-matched sample within 2.$'$0 of Kron3. Blue point indicates expected variation due to variation in N abundance.
}
\label{CN_spread}
\end{figure}

\begin{figure*}
\centering
\includegraphics[width=\columnwidth]{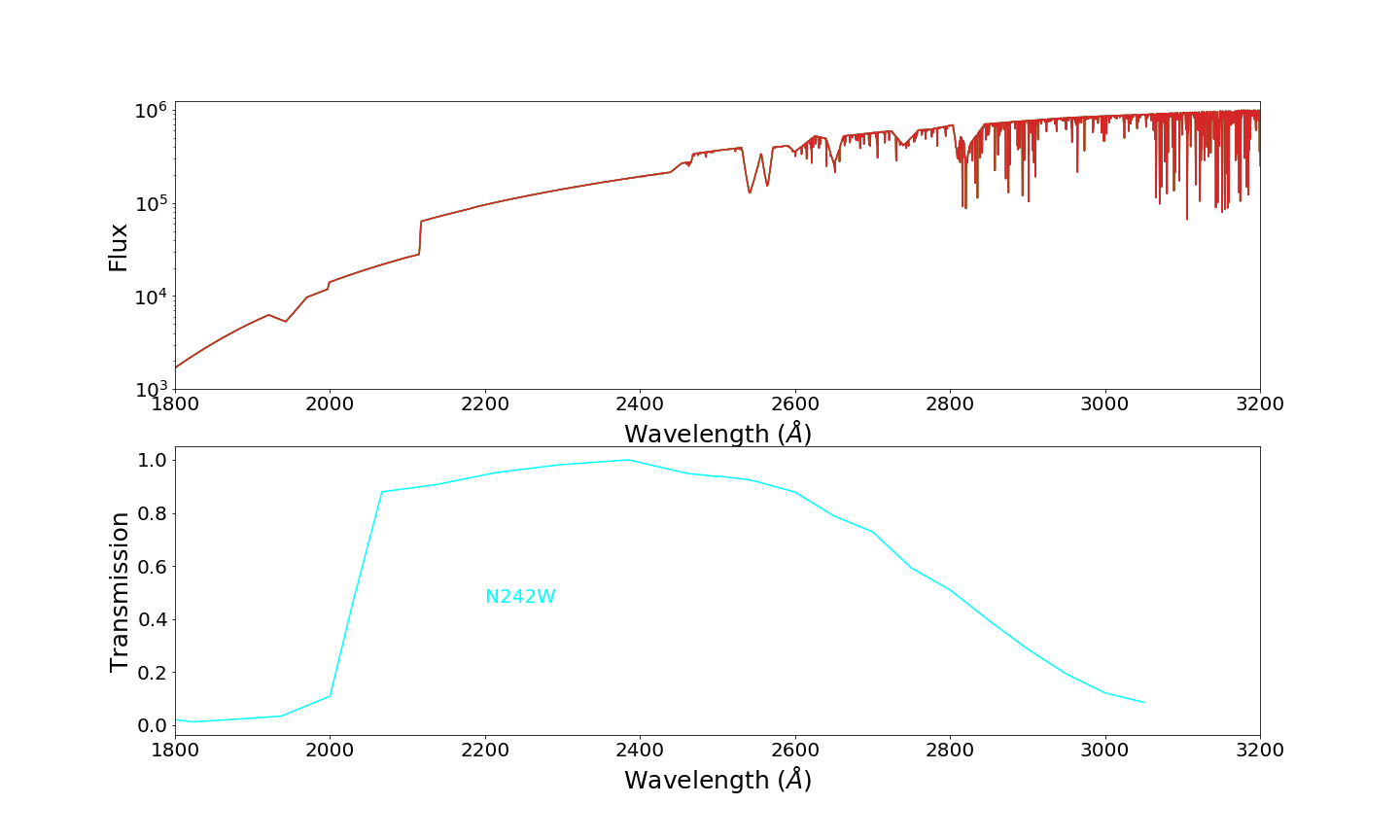} \includegraphics[width=\columnwidth]{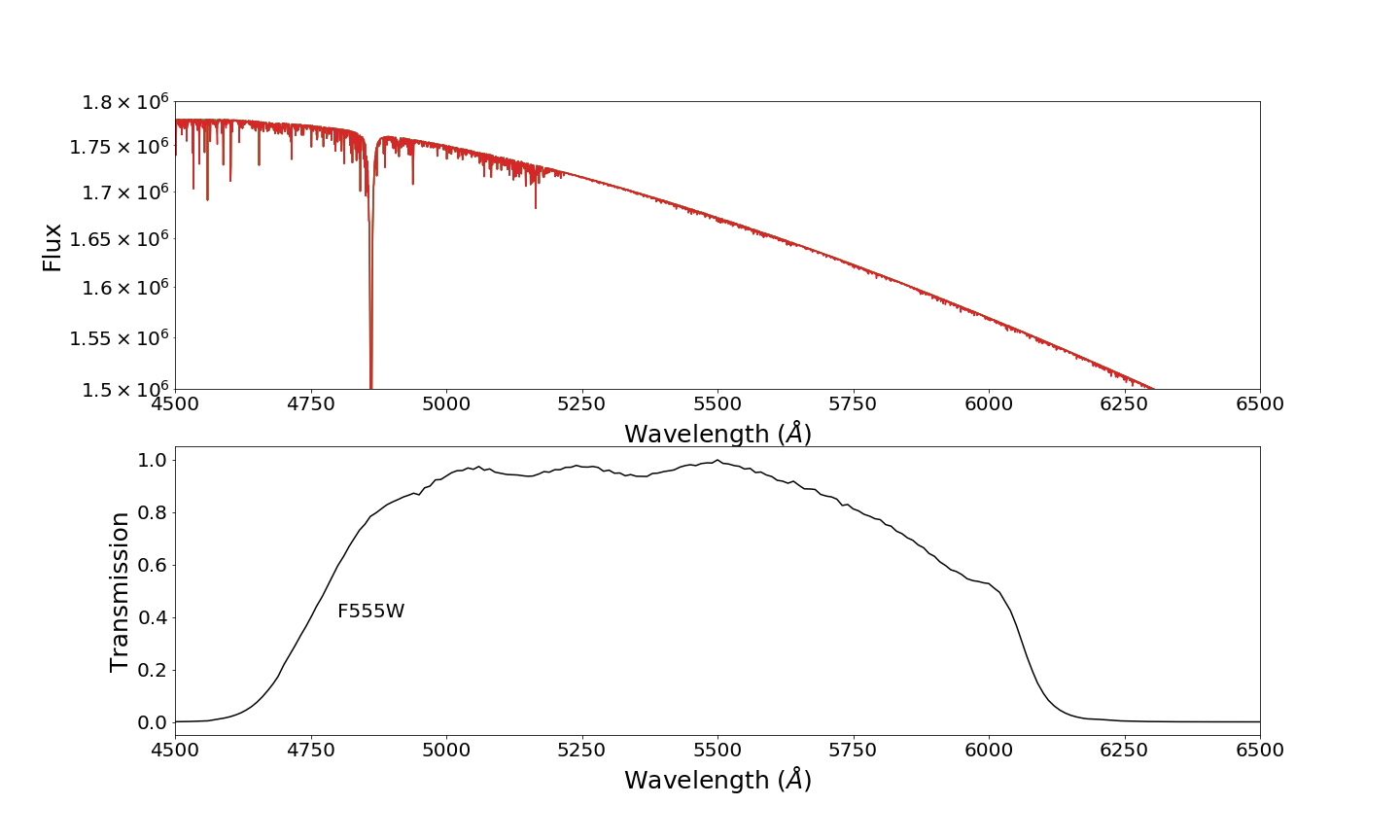}
\caption{Model generated synthetic spectra for different N abundances [N$/$Fe] = -1.0 to +1.50 dex (marked in different colours) are presented in the upper panels, temperature 5250 K and log(g)=2.5. In the lower panel the plot shows response curves for N242W and F555W pass bands.}
\label{N_spec}
\end{figure*}

\subsubsection{C-N-O variations}

To check the effect of CNO variation in Kron 3, we first generated synthetic spectra for RC stars for a fixed temperature (5250K) and logg (2.5) but for different elemental abundances. The values of temperature and logg values are obtained from Padova theoretical isochrone model.  
We have employed the one-dimensional local thermodynamic equilibrium LTE stellar atmospheric models (ATLAS9; \citealt{castellikurucz}) and the spectral synthesis code TURBOSPECTRUM \citep{alvarezplez1998} to construct synthetic spectra. Version 12 of the turbo-spectrum code for spectrum synthesis and abundance estimates was used for the analysis. We have generated the spectra by varying the abundances of C and N for a spectral range from 1500 to 10500 angstrom (for the adopted stellar model) in order to measure the variation in flux. Taking solar abundances as the reference, we have varied [N$/$Fe] between -1.0 to +1.50 dex in steps of 0.50, whereas [C$/$Fe] was varied between 0.0 and -1.0 dex following the typical abundance range for globular clusters of a given metallicity. 
After generating the spectra, we have convolved them with filters' response curves using VOSA module to calculated the expected flux in each filter for different elemental abundances. We converted the model flux to expected observed flux using scaling factor ($M_d = (R_c/D)^2$). 
The equation for flux to magnitude conversion has been used to convert the expected flux to magnitude unit. The zero point flux values for the corresponding filter system. As the converted magnitudes are Vega systems, we used the Vega to AB conversion factors for corresponding filters from MESA website. The converted AB magnitude then over plotted on observed (NUV – V) vs NUV CMD after correcting for extinction. 

Figure \ref{CN_spread} shows the UV-optical CMD of the cluster where RC stars are highlighted with red points. All the points for the range of [N$/$Fe] abundances (-1.0 to +1.50 dex) and for the range for [C$/$Fe] abundances (0.0 to -1.0 dex) merge together in a single blue point in the Figure \ref{CN_spread}. The Figure suggest that C and N variations are unable to produce the observed spread. In the Figure \ref{N_spec}, we have plotted synthetic spectra for different N abundances in the spectral region of F555W and N242W filter systems. The Figure clearly shows that there is almost no variation in the flux for variation in N abundances. As shown in figure 32 of \citealp{milone2012_47tuc}, N variation affects are seen more near $\sim$3320-3580 angstrom (NH band), $\sim$3700-3880 angstrom (CN band) and does not affect much in other wavelength region. Our model generated synthetic spectra also show flux variations in the aforementioned wavelength range for NH and CN molecular band (not shown here) but does not show enough variation in the wavelength range for N242W filter. 
From our analysis, it is clear that the N242W filter is insensitive to variations in either C or N, and as such, we can rule out such variations as significant contributions to the observed spread in the RC stars in the NUV.


\subsubsection{Spread in initial helium abundances}

\begin{figure*}
\centering
\includegraphics[width=2\columnwidth]{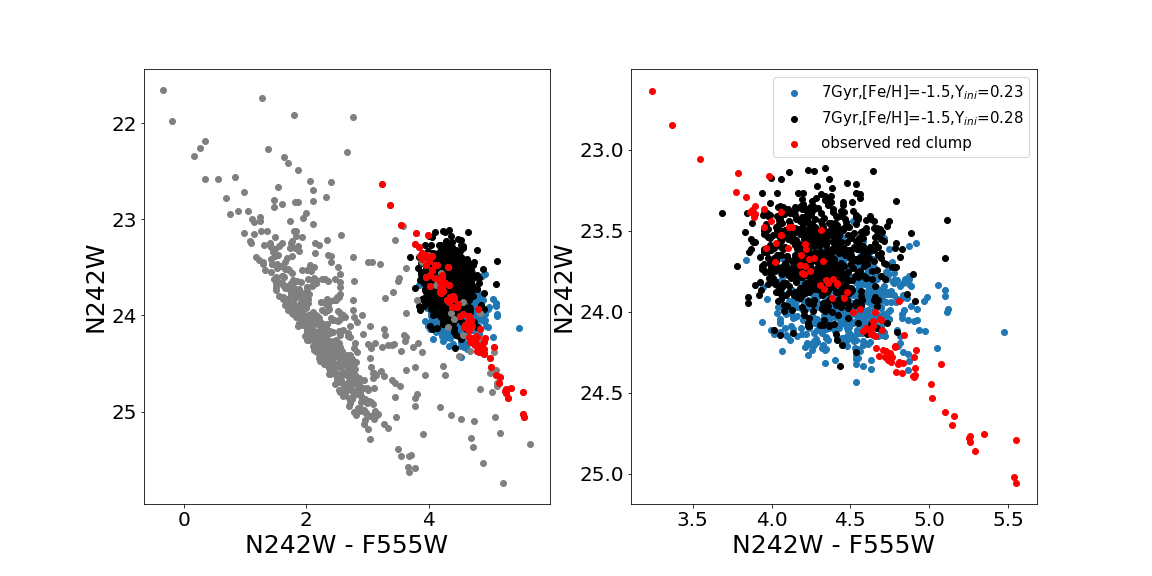}
\caption{NUV vs ($NUV-V$) CMD of the HLA-UVIT cross-matched sample within 2.$'$0 of Kron3. Blue and black points indicate expected spread in RC due to different initial helium abundance (Y$_{ini}$) of 0.23 and 0.28, respectively, but for a fixed age (7 Gyr) and metallicity [Fe/H]=$-$1.5.}
\label{He_7gyr}
\end{figure*}

\citet{chul2013, chul2017} showed that implications and prospects for the helium-enhanced populations in relation to the second-generation populations found in the Milky Way GCs using Yonsei Evolutionary Population Synthesis (YEPS) model. 
The model provides us Yonsei–Yale (${\rm Y}^2$) stellar evolutionary tracks and BaSel 3.1 flux libraries. To construct synthetic CMD for Helium-enhanced red HB stellar populations , we have used ${\rm Y}^2$ stellar libraries with enhanced initial helium abundances (Y$_{ini}$) (Lee et al.2015). We choose two values for Y$_{ini}$ as 0.23 and 0.28, at fixed Z value of 0.001 ([Fe/H]=-1.5 dex) and age of 7 Gyr. We have also considered $\alpha$-enhancement [$\alpha$/Fe]=0.3, under the $\alpha$-elements mixture of Kim et al.(2002). We chose Salpeter's IMF for the model \citep{salpeter1955}. 
The left panel of Figure \ref{He_7gyr} shows that synthetic CMDs for two different Y$_{ini}$ values 0.23 (blue) and 0.28 (black), over plotted on the observed NUV-optical CMD where observed RC stars are highlighted in red. The right hand panel shows the zoomed version of RC or red HB population. The observation photometric errors have been included to generate the synthetic CMD. We notice that the spread in Y$_{ini}$ can produce NUV bright stars and is able to explain a large fraction of observed extension in the RC. Therefore, we can suggest that initial variation in helium is one of the reason behind the observed spread in RC population.

\begin{figure}
\centering
\includegraphics[width=\columnwidth]{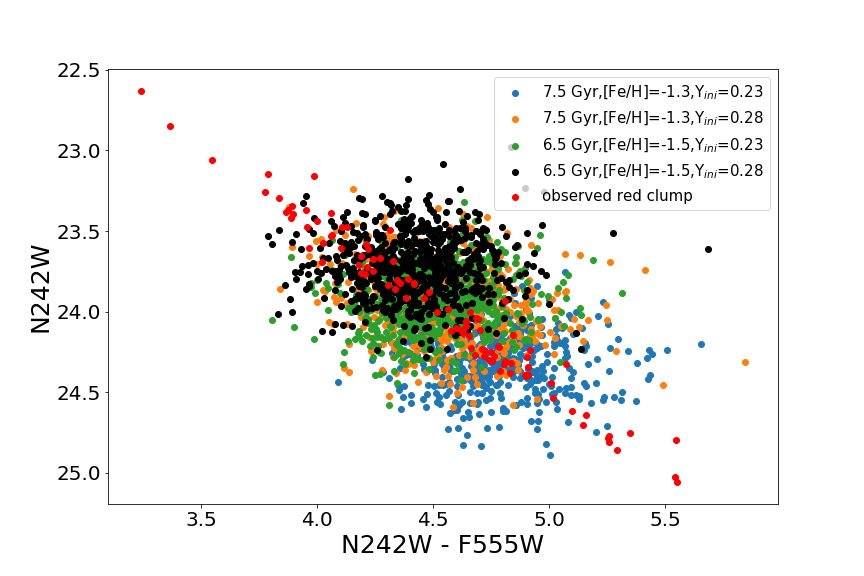}
\caption{The same as Fig. \ref{He_7gyr} but for different age and metallicity combinations.}
\label{He_6&8gyr}
\end{figure}

Now, to check if a small spread in age and metallicity along with the variation in Y$_{ini}$ can accout for the full spread observed in RC, we generated synthetic CMD for HB stars for two different combinations of age and metallicity along with initial helium variation of Y$_{ini}$=0.23 \& 0.28: (1) metal-poor ([Fe/H]=-1.5 dex) and slightly younger (6.5 Gyr); (2) metal-rich ([Fe/H]=-1.3 dex) and slightly older age (7.5 Gyr). 
In Figure \ref{He_6&8gyr}, We show synthetic CMDs for these combinations on observed RC stars, which clearly suggests that a spread in helium abundance along with a small spread in age and metallicity able to produce observed extension in the red clump. 
So, we can suggests that the cluster hosts stars with a small spread in age and metallicity along with a variation in initial helium abundances.

\section{Discussion}

Using both the UVIT and $Gaia$-DR2 data sets, it is found that the distribution of stellar density around the centre of Kron 3 matches with the field star distribution at a distance of $\sim$2$'$, suggesting the radius of Kron 3 to be $\sim$2$'$ ($\sim$35 pc). This estimate is slightly larger than a previous estimate of 1.$'$7 by \cite{bica2008} and lower compared to an estimate (2.$'$4) by \cite{rich1984}. Our result does not support the claim by \cite{alca1996}, where the authors suggested a radius of  6$'$, as we do not see any variation in the star count beyond 2$'$ in both the optical and NUV. 

In general, the presence multiple stellar populations can be attributed to large variations in the elemental abundances and have been identified in Galactic globular clusters, which were once thought to be chemically homogeneous. Similar variations have recently been found in the only $'$classical$'$ globular cluster NGC 121 ($\sim$10 Gyr old) and a few intermediate-age ($\sim$6-8 Gyr old) clusters in the SMC (Lindsay 1, NGC 416, NGC 339; \citealp{neider2017a,neider2017b}).  
The authors used multi-band photometric survey data obtained with the WFC3/UVIS instrument on board the HST and constructed pseudo-colour magnitude diagrams in order to distinguish the different populations present in the cluster, where pseudo-colour is defined as C$_{F336W,F438W,F343N}$ = (F336W $-$ F438W) $-$ (F438W $-$ F343N). If C and N abundance variations are present within a given cluster, the use of the pseudo-colour-magnitude diagram acts to separate the different RGB populations in a way that standard colour-magnitude diagrams are unable to do.
The fraction of N enriched and C poor stars found in NGC 121, NGC416, NGC 339 and Lindsay 1 are $\sim$32, 45, 25 and 36 per cent, respectively \citep{neider2017a,neider2017b}. These fractions are found to be lower than the average value for Galactic globular clusters.
\cite{holly2017,holly2018} provided spectroscopic evidence of N-enriched stars in the RGB population of Lindsay 1 and Kron 3. All the clusters mentioned above are equally massive with a mass range of 1-2$\times$10$^5$ M$_\odot$ and located in different regions of the SMC, hence mass and environmental effects are not key for the presence of chemical enrichment in these cluster \citep{marto2017}. The morphology of the RC in these clusters has not been studied so far; here we present the first detailed analysis of the RC in Kron 3, one of the youngest clusters where elemental abundance variations identified.

UVIT images in the NUV allow us to detect RC stars in Kron 3 with good photometry. We demonstrate that Kron 3 exhibits an extended RC in UV-optical CMDs. This extension could be due to photometric uncertainties, PSF variation and/or differential reddening across the cluster region, the influence of field stars, an age spread, a metallicity gradient, variation in light elemental (C-N-O) abundances, variation initial helium abundances or to a variable stellar mass loss. We investigated these aspects to explain the extension of the RC. We noticed that photometric uncertainties, differential reddening, PSF variation and field stars contamination are not able to curtail the RC extension, suggesting that it is likely an intrinsic property of the cluster.

We found that variable stellar mass loss produce a small spread for the RC stars, not enough to explain the observed spread in RC. So, variable mass loss may not be the sole reason behind the RC spread. 
We have also examined whether the extension is due to multiple stellar populations arising out of age or metallicity variations. With the help of isochrones, we found that an age spread is unable to fit the extended RC distribution. Theoretical isochrones revealed that the colour of RC stars is independent of age. We suggest that the cluster has a small range in age, though we are unable to quantify.

We used three methods to check whether a range in metallicity can reproduce the observations.
These are: (i) isochrones, (ii) colour from model spectra and (iii) SEDs. We found that isochrones of different metallicities for an age of 7 Gyr are able to explain the extended RC. 
Kron 3 shows a range of metallicities from Z=0.0002 to 0.002. The lower metallicity RC stars appear hotter and brighter in UV-optical CMDs, while higher metallicity RC stars appear relatively fainter and cooler. 
The colours estimated from model spectra with temperature and log(g) corresponding to the RC stars and a metallicity within the range above, were found to remarkably fit the RC distribution in the observed CMDs. The NUV passband is found to be sensitive to metallicity because of the spectral features that fall within it. Thus, this band is suitable to study metallicity variations among RC stars. 
To reproduce the range in temperature and metallicity of the RC stars we constructed SEDs and fitted them with model spectra. We used data that cover a large range of wavelengths, from the NUV to the near-IR. The best-fit model spectra suggest that RC stars span a large metallicity range, between Z=0.00006 and solar, and have effective temperatures between 5250 and 5500 K. We found that a reduced metallicity range is obtained by fixing log(g) to a typical value for RC stars. From the fitted spectrum, we have also found the values of luminosity and radius of these stars lie in the range $\sim$60-90 L$_{\odot}$ and $\sim$8-11 R$_{\odot}$, respectively, which are consistent with those expected of RC stars \citep{gallenne2018, wan2015}.

All three methods suggest that the RC stars span a wide range in metallicity and Kron 3 is probably the first cluster where this has been quantified. A spectroscopic abundance study of RC stars in Kron 3 has not been performed so far. \cite{dias2010} estimated the metallicity of Kron 3 as Z=0.0002, which falls within the metallicity range found in this study. Our study supports that Kron 3 is a relatively young cluster of age 7 Gyr with respect to ancient GCs hosting multiple stellar populations.

Metallicity gradient among the stars suggest that variation in the strength of absorption lines due to variation by Mg, C, Fe II, Iron peak elements etc in the NUV region causes the extended RC. Whereas, the variation in the light elemental abundance (He, C, N) can also cause variation in molecular absorption line as well as in continuum.  
Therefore, we have also examined whether effect of variation in elemental abundances in RC stars. We generate synthetic CMDs for C and N abundances for a fixed temperature (5250K), logg (2.5), metallicity (Z=0.001) and convolve with filter response curve to observed spread in UV-optical CMD. We found almost no variation in the CMD, suggest that elemental abundance variation is not the cause of observed spread in RC. As, previous study by \citealp{holly2018} suggested presence of C and N variations in the RGB stars but we do not see any signature of that in RC stars using NUV filter, so we also suggest that NUV filter might not be sensitive to the elemental abundance variation.

We also tried to check if spread in initial helium abundance (Y$_{ini}$) can mimic the observed extension in the RC, as previous study by \citealp{chantereau2019} showed the presence of variation in initial helium abundances in the RC distribution of a very similar (having similar age, metallicity and reddening) SMC cluster Lindsay 1. We used alpha enhanced ([$\alpha$/Fe]) ${\rm Y}^2$ isochrone model to generate synthetic CMD of RC stars for two different initial value of helium (0.23 and 0.28) keeping age and metallicity constant as average value for the cluster 7 Gyr and Z=0.001 (or [Fe/H]=$-$1.5), respectively. Distribution of synthetic RC stars clearly suggest that variation in Y$_{ini}$ can mimic the observed spread in RC to a large extend. We have further found that combination of a small spread in metallicity by 0.2 dex, a small age spread of 6.5-7.5 Gyr along with variation in Y$_{ini}$ able to produce the observed spread very nicely. 

In optical CMD, RC stars have an almost constant luminosity throughout their lifetime but T$_{eff}$ can vary depending on their metallicities and initial masses \citep{girardi2016, choi2016}. 
For a constant stellar mass, the luminosity of an RC star varies with the metallicity. Stars with a lower metallicity will have higher luminosity and vice versa \citep{girardi2016}. In our study, we found that combination effect of variation in Y$_{ini}$ and a small spread in metallicity can cause a large spread in NUV luminosity of RC stars which show a constant luminosity and very compact distribution in the optical CMD.  
\cite{mohammed2019} observed NUV bright RC stars in the solar neighborhood and found that the absolute (NUV$-$G) colour is strongly correlated with spectroscopic metallicities, whereas the absolute ($G_{\rm{BP}}-G_{\rm{RP}}$) colour has a weak correlation with metallicity and a large scatter. Bluer stars tend to be hotter and have a lower metallicity and vice versa. This relation can be used to obtain photometric metallicities from other stars in the same CMD space as RC candidates using only their UV-optical colour. They also found that metallicity vs absolute (NUV$-$G) colour in the MIST models is only weakly dependent on the initial mass and age during the RC phase of evolution. Though they have not taken care of the effect of initial helium variation among the stars. 
Our study of Kron 3 suggests that the NUV$-$optical colour can be used as a tool to investigate the presence of multiple stellar populations  due to initial helium variation and also for metallicity spread within clusters.

After taking into account the effect helium variation, we exclude the possibility of a large spread in metallicity. We suggests that the observed spread in RC is an combined effect of photometric error, variation in Y$_{ini}$, variable mass loss, a small spread in metallicity and age. We are not quantifying the value of initial helium spread, age and metallicity spread as there will be an uncertainties in those values due to relatively large photometric error in fainter end of NUV magnitude and to degeneracy between metallicity and helium spread. 
The recent studies by \citealp{marino2019,milone2020_type2, milone2020_wcen} found the presence of Type II GCs in the Milky Way using universal chromosome map (ChM). All the GCs shows both 1G and 2G population, suggesting the multiple populations in them. The universal ChM helps to reveal the internal variation of elemental abundances or metallicity ($>$0.1) or helium within the 1G stars of Type II GCs, whereas Type I GCs do not show any any internal variation. We are unable to separate two generation of stars, rather found a continuous spread of different population of stars, which is explained by both variation in metallicity ($\sim$0.2 dex) and helium. Hence, we suggest that a more details study is required using HST/UVIS filters, where we can generate universal ChM to check if Kron 3 could be a probable candidate of Type II GCs in the Magellanic Clouds. A spectroscopic follow-up study of RC stars will also be helpful to check if the metallicity spread is there or not.

\section{Summary}

 We summarize the results of our study of the Kron 3 cluster in the SMC as follows. \ 
 
We present the analysis of UVIT-HST-{\it Gaia}-VISTA data for the intermediate-age cluster Kron 3 in the SMC.  For the first time, we report the identification of NUV bright RC stars and the extension of the RC in the CMD. This study thus demonstrates the power of UVIT-HST-{\it Gaia}-VISTA combination to study clusters in the Magellanic Clouds. 
We take advantage of the high spatial resolution of the HST in the central region of the cluster and of the wide-area coverage of Gaia in combination with the UVIT data. 
We find that the extended RC is an intrinsic property of the cluster and that it is not due to the influence of field stars. We estimated the radius of the cluster as 2.$'$0 (35 pc) from the UVIT and {\it Gaia} DR2 data.
 
We suggest that the cluster exhibits multiple stellar populations with a small range in age (6.5-7.5 Gyr) and metallicity (0.2 dex) along with a variation in initial helium abundance Y$_{ini}$=0.23 to 0.28. 
We suggest that NUV filter can be used as a tool to investigate the presence variation in initial helium abundance and metallicity spread within a cluster.  
A spectroscopic follow-up study of RC stars are suggested to check if Kron 3 is a probable candidate of Type II GCs in the SMC.

\section*{Acknowledgements}

This research was supported by a DST-DAAD exchange grant (REMAP) between the Indian Institute of Astrophysics (IIA, Bangalore) and the Leibniz Institute for Astrophysics Potsdam (AIP, Potsdam).
This paper makes use of observations collected at the European Organisation for Astronomical Research in the Southern Hemisphere under ESO programme 179.B-2003. We thank the CASU and the WFAU in Edinburgh for providing calibrated data products under the support of the Science and Technology Facility Council (STFC) in the UK. 
M.-R.L. Cioni and Cameron P. M. Bell acknowledge support from the European Research Council (ERC) under the European Union's Horizon 2020 research and innovation programme (grant agreement no. 682115). S. Subramanian acknowledges support from the Science and Engineering Research Board (SERB), India through Ramanujan Fellowship. We thank the referee for valuable suggestions to help in improving the manuscript. 
P. K. Nayak  thank  William Chantereau and Nate Bastian for valuable discussion with them.




\bibliographystyle{mnras}
\bibliography{kron3} 



\appendix

\section{Model fitted SEDs of RC stars}


\begin{figure*}
\centering
{\includegraphics[width=0.74\columnwidth]{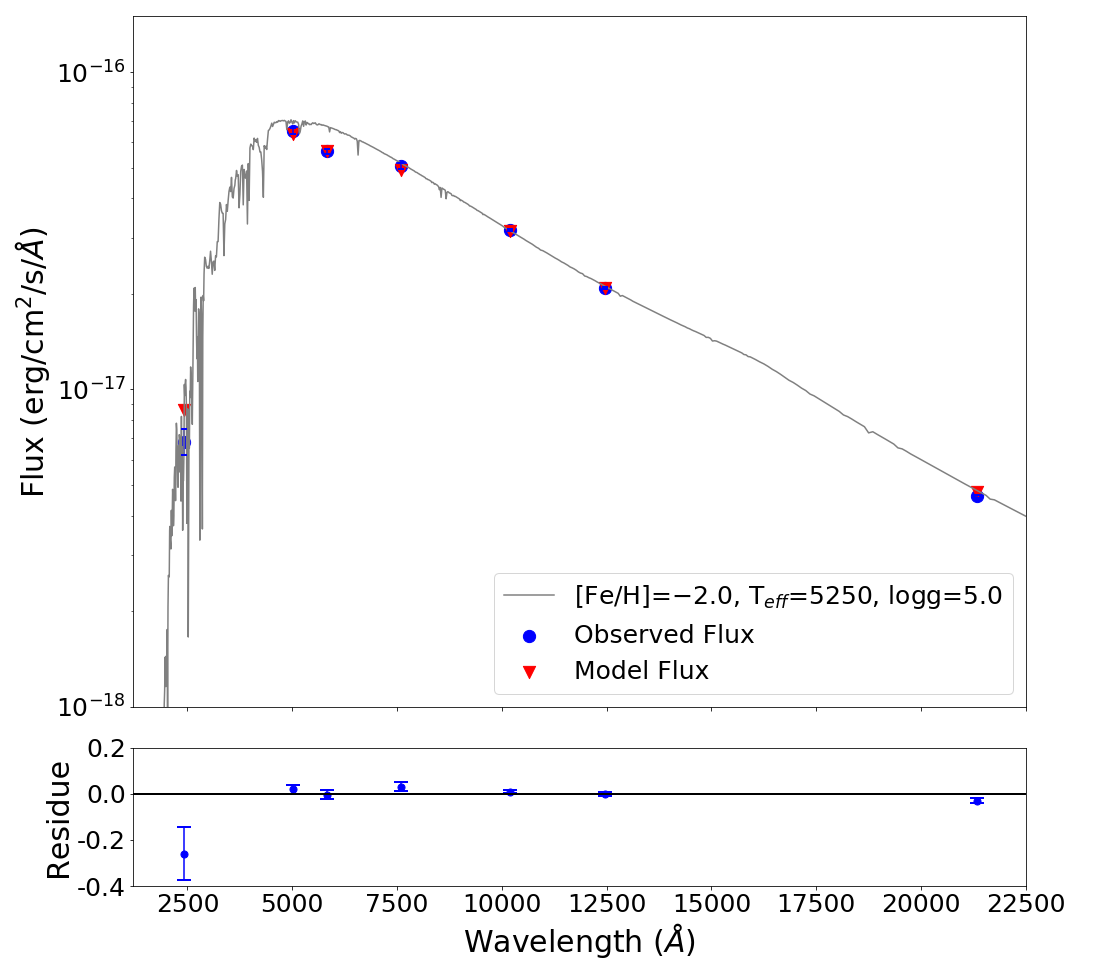}} {\includegraphics[width=0.74\columnwidth]{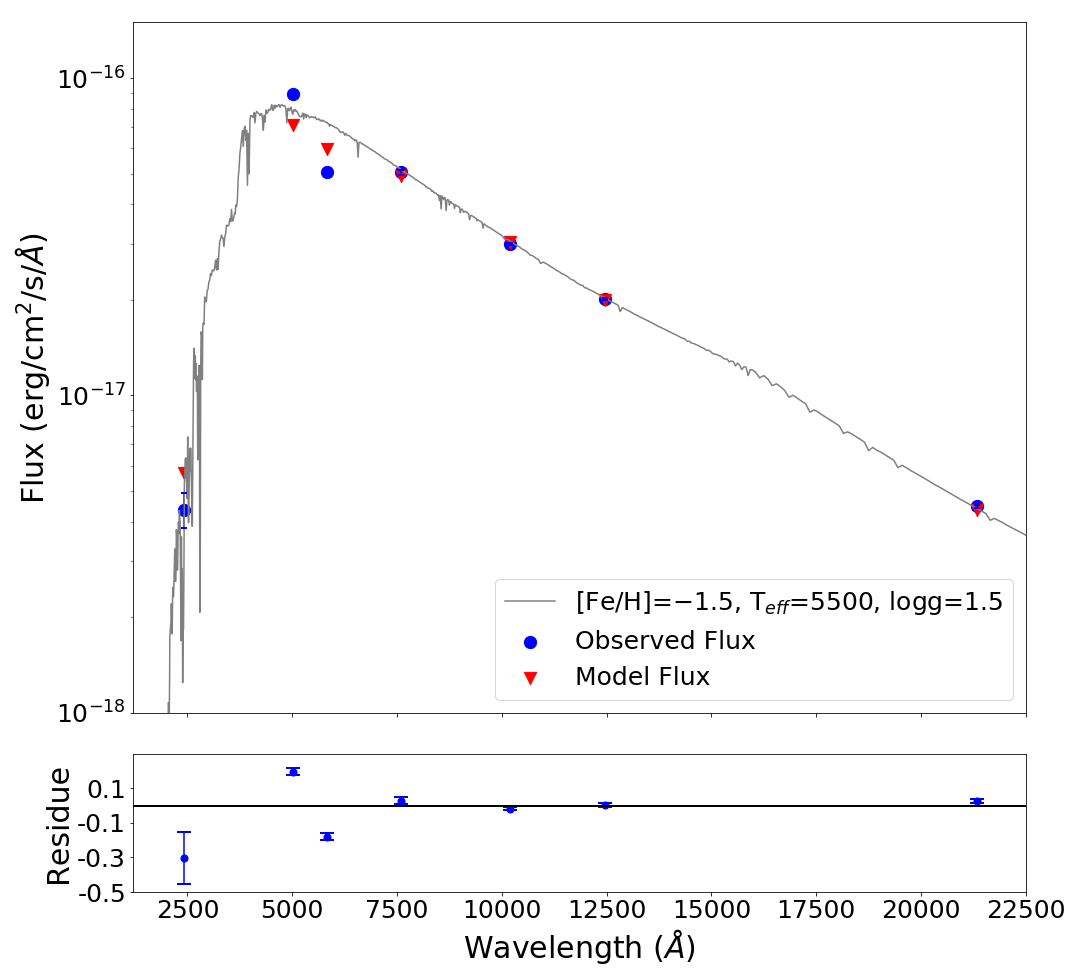}} \ 
{\includegraphics[width=0.74\columnwidth]{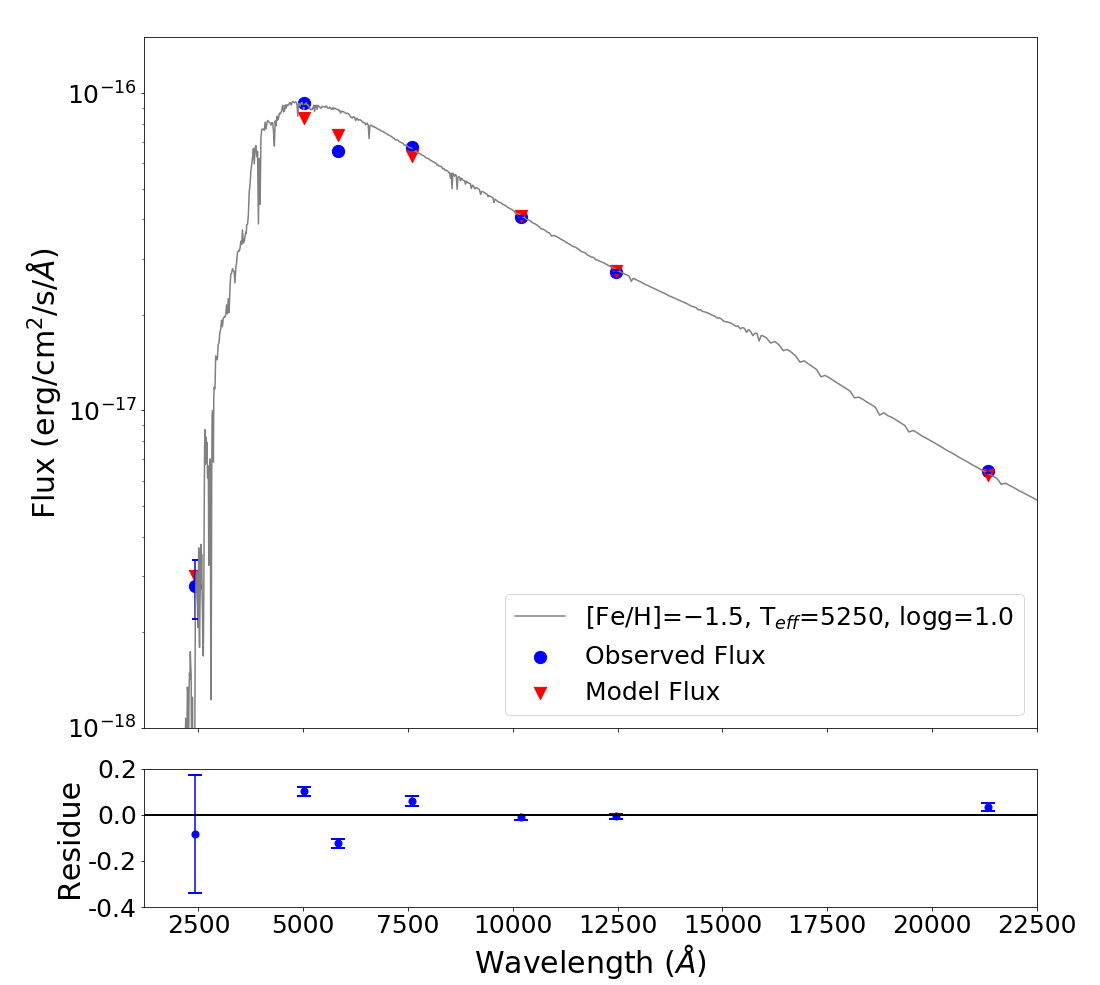}} {\includegraphics[width=0.74\columnwidth]{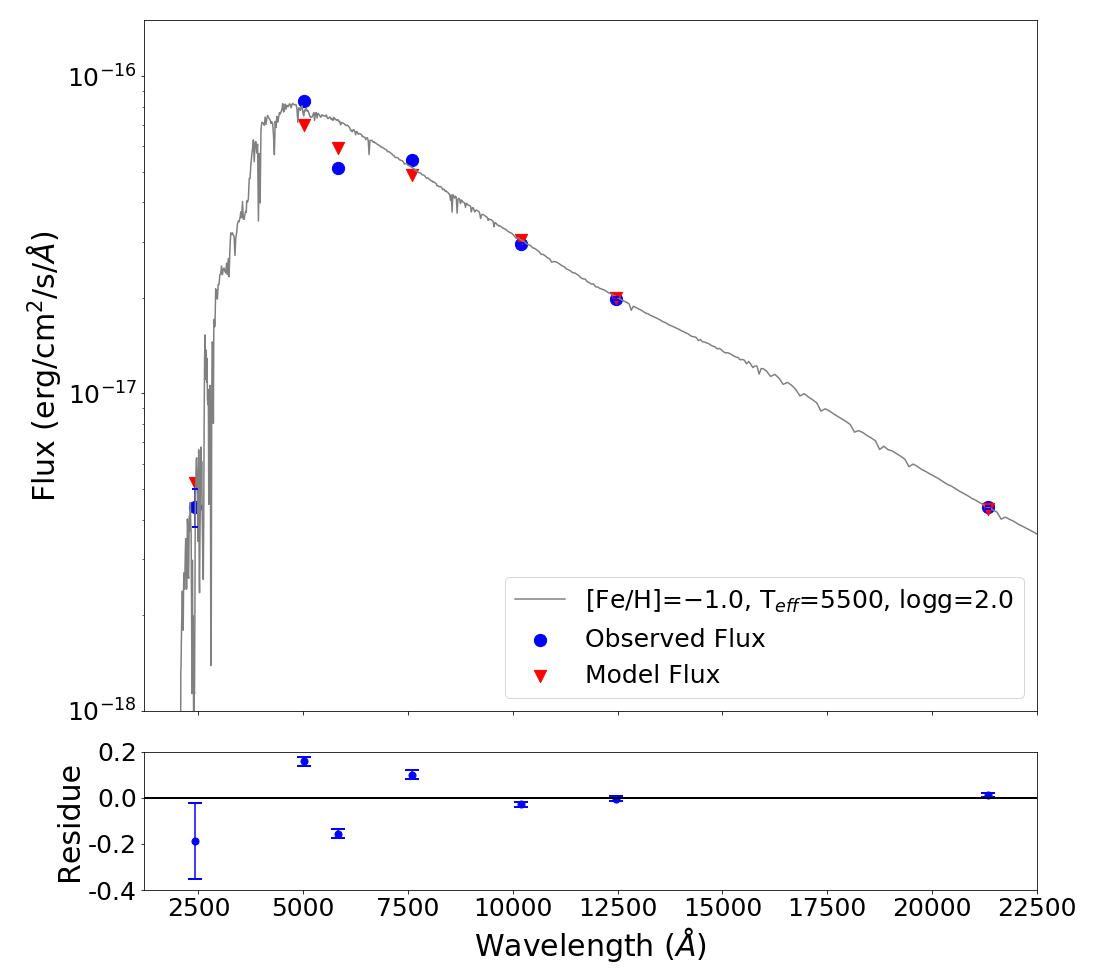}} \
{\includegraphics[width=0.74\columnwidth]{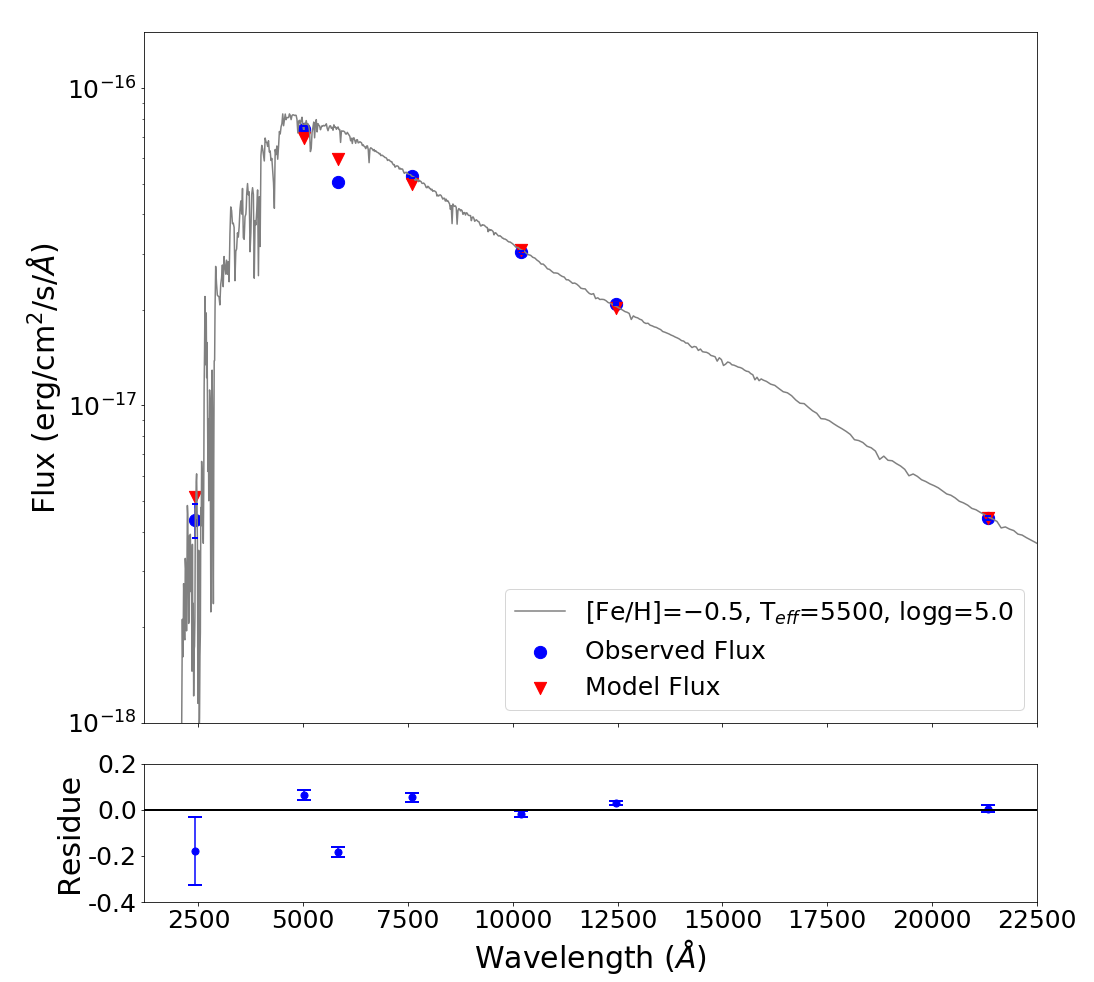}} {\includegraphics[width=0.74\columnwidth]{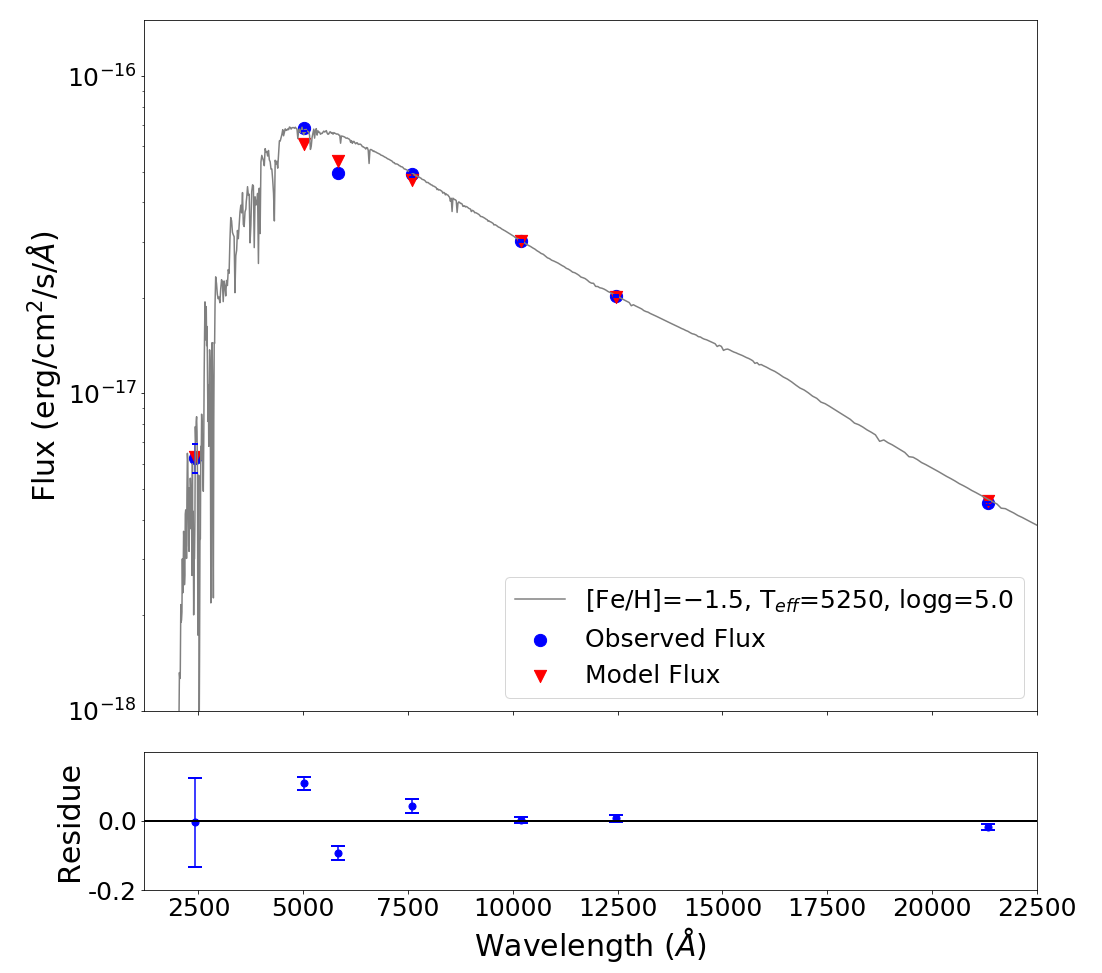}} \ 
{\includegraphics[width=0.74\columnwidth]{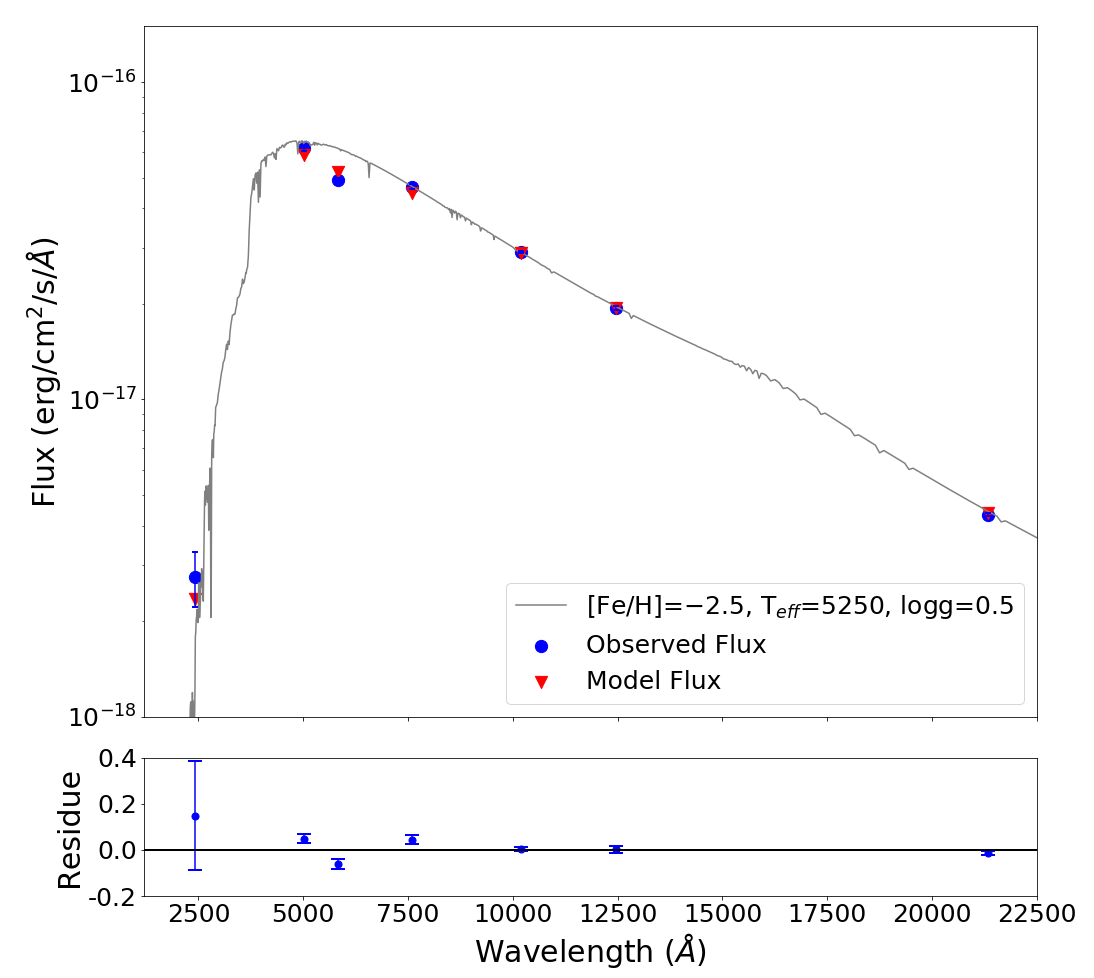}} {\includegraphics[width=0.74\columnwidth]{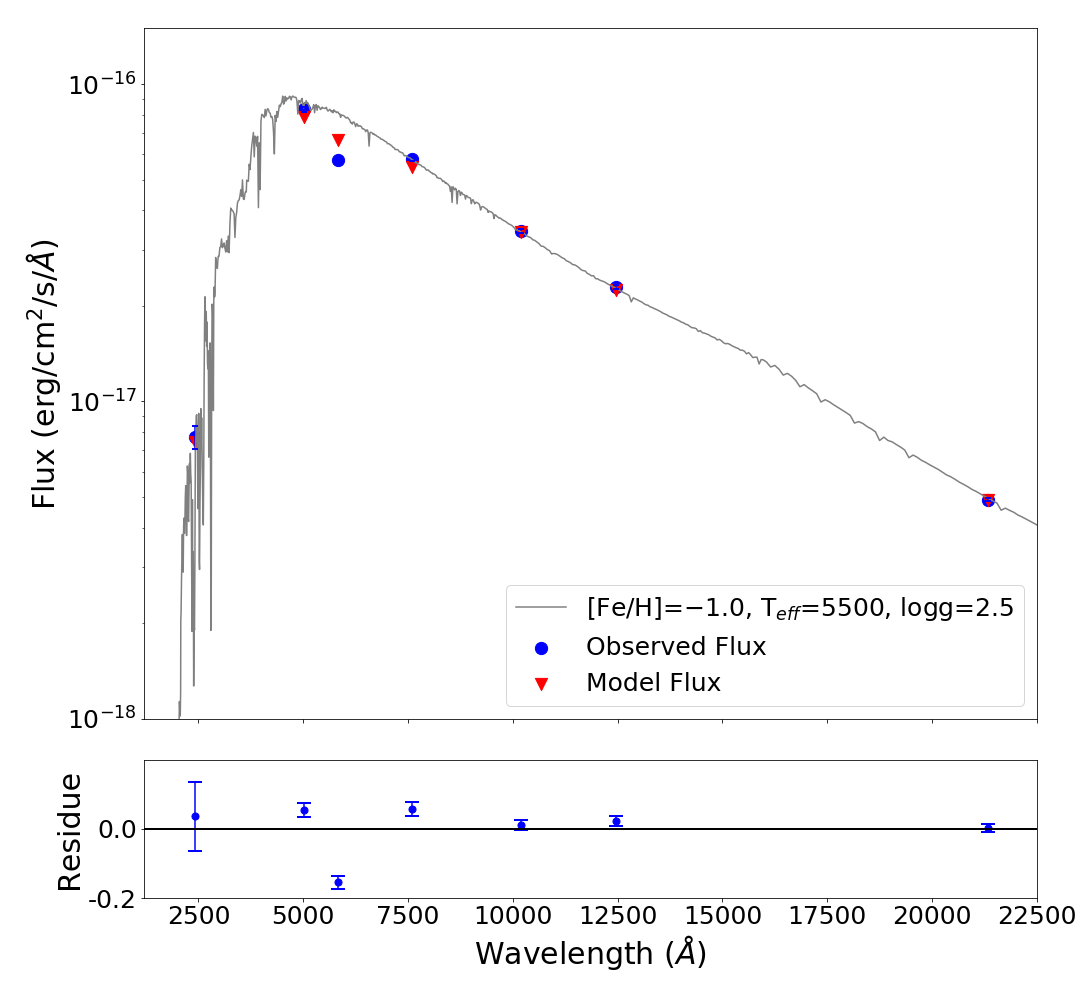}}
\caption{SEDs of eight RC stars are presented in UVIT, Gaia DR2 and VMC data. 
The blue and red points represent the observed and expected fluxes, respectively in different passbands. 
}
\label{sed3}
\end{figure*}

\begin{figure*}
\centering
{\includegraphics[width=0.74\columnwidth]{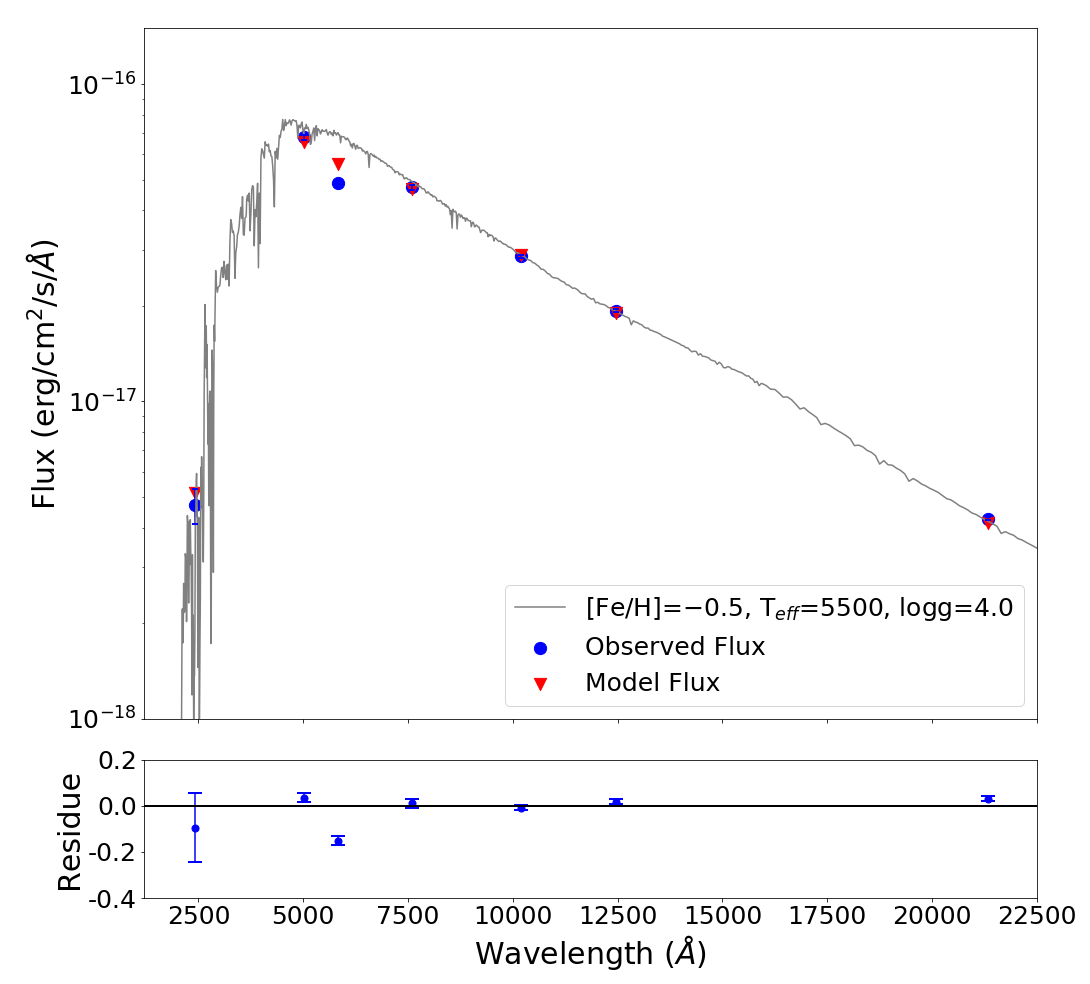}} {\includegraphics[width=0.74\columnwidth]{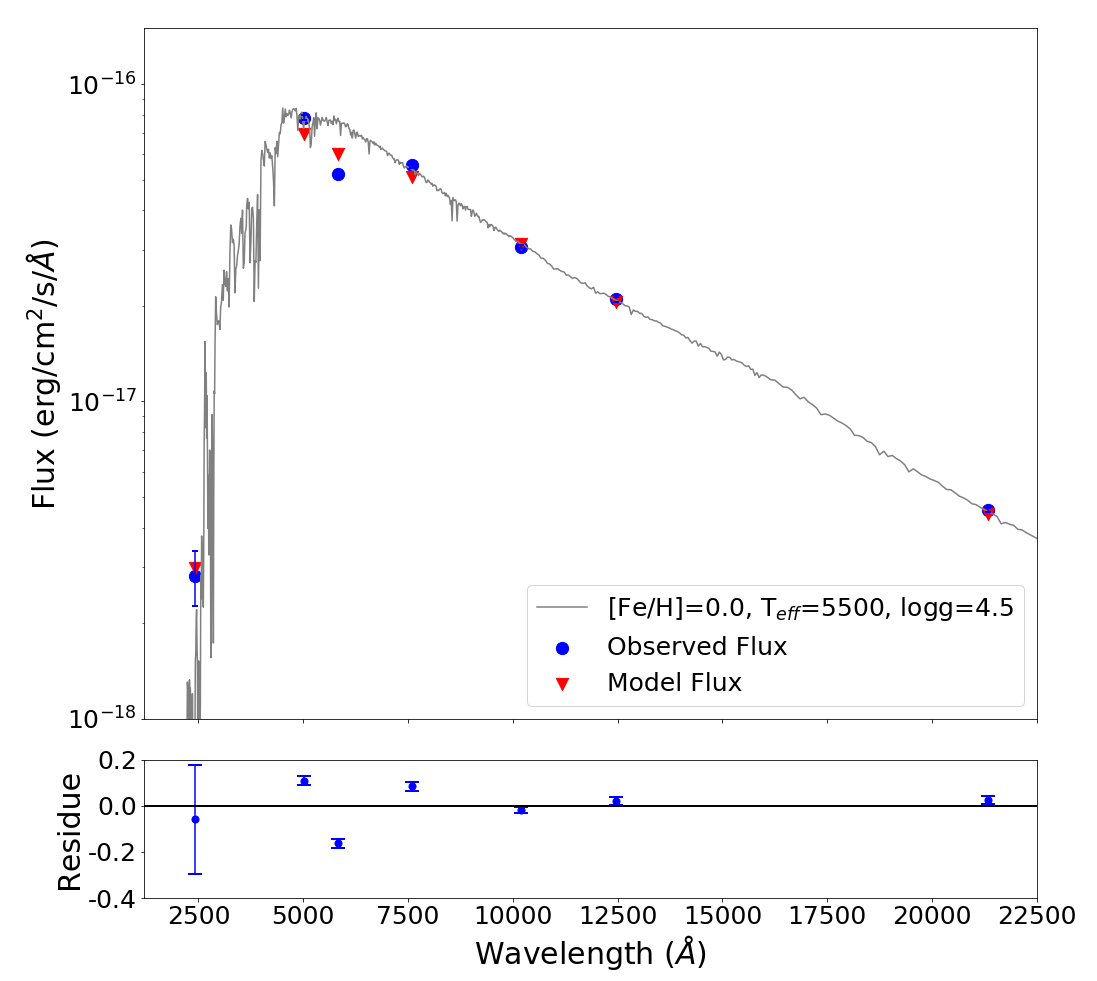}} \ 
{\includegraphics[width=0.74\columnwidth]{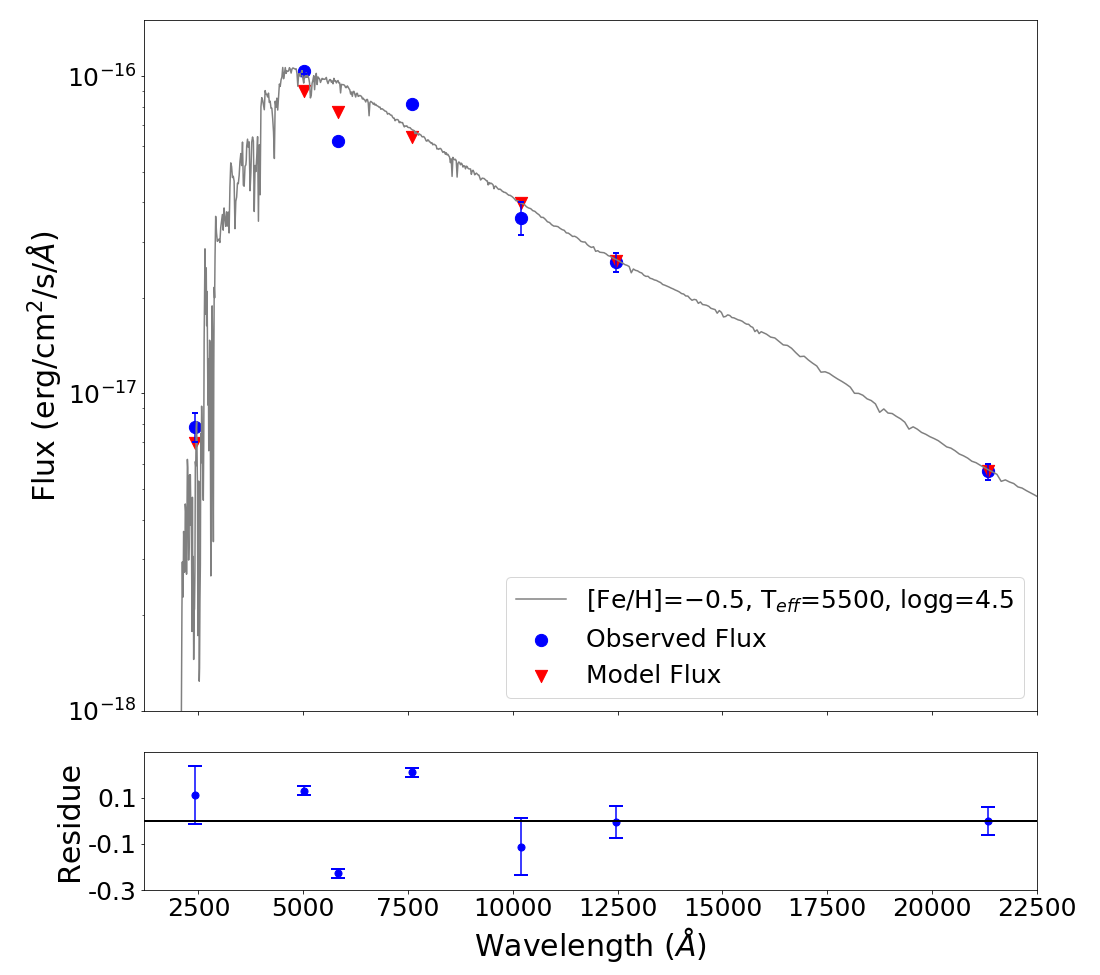}} {\includegraphics[width=0.74\columnwidth]{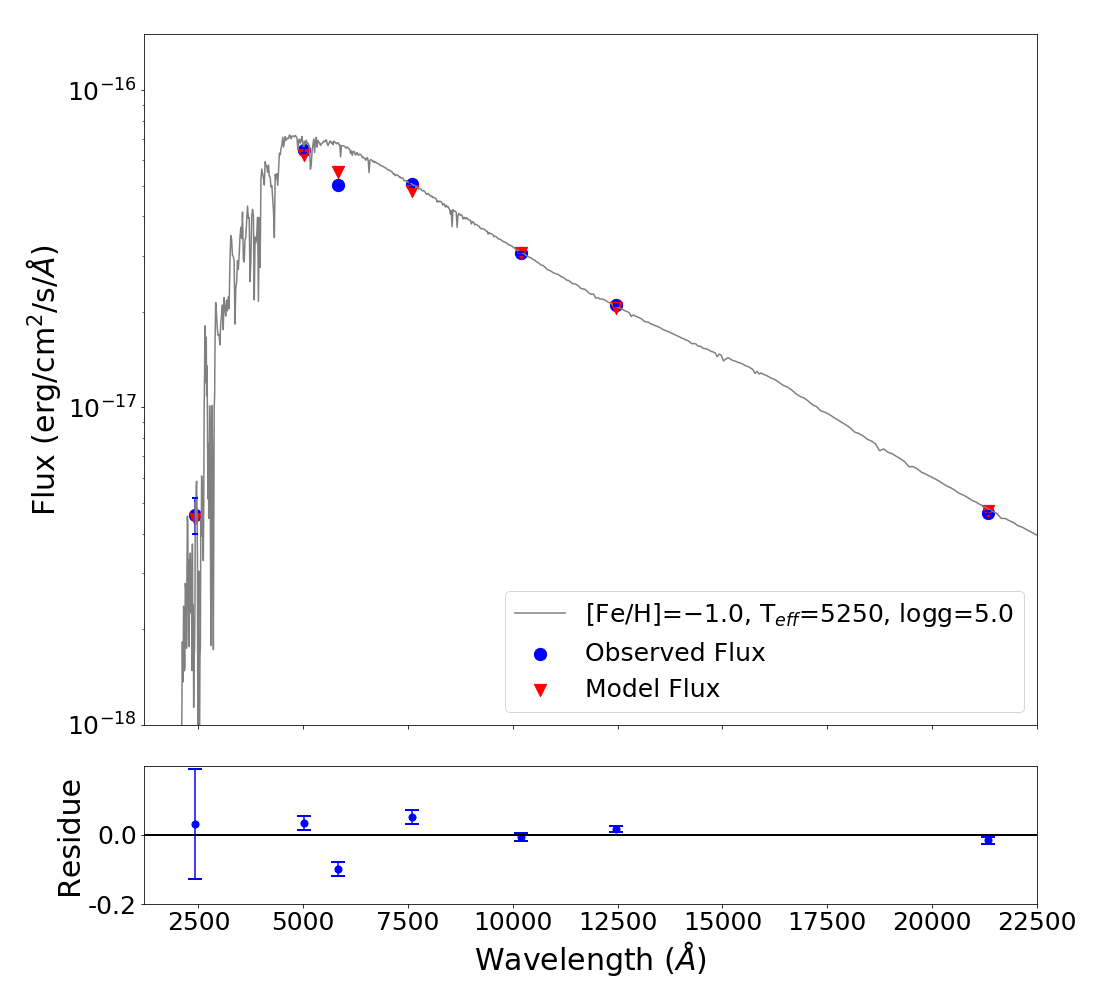}} \
{\includegraphics[width=0.74\columnwidth]{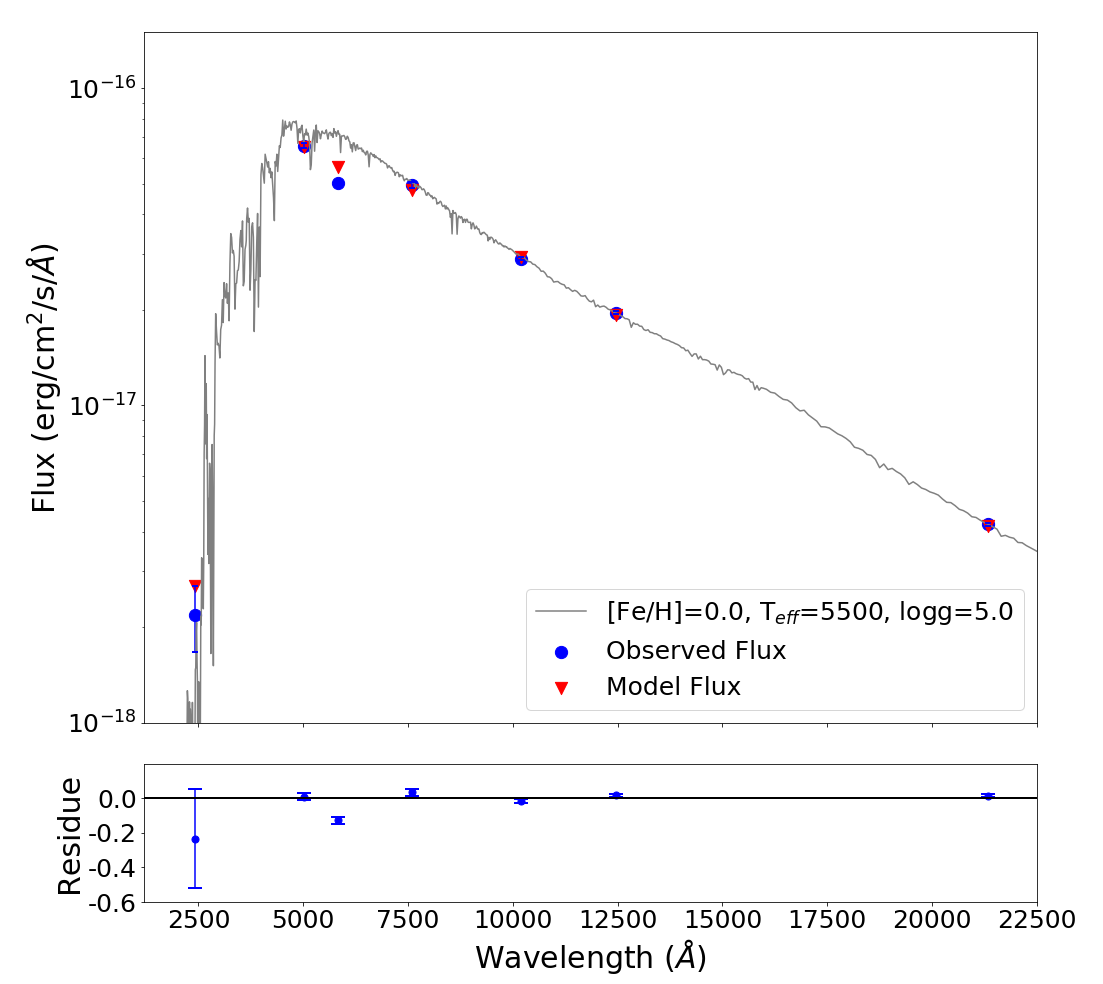}} {\includegraphics[width=0.74\columnwidth]{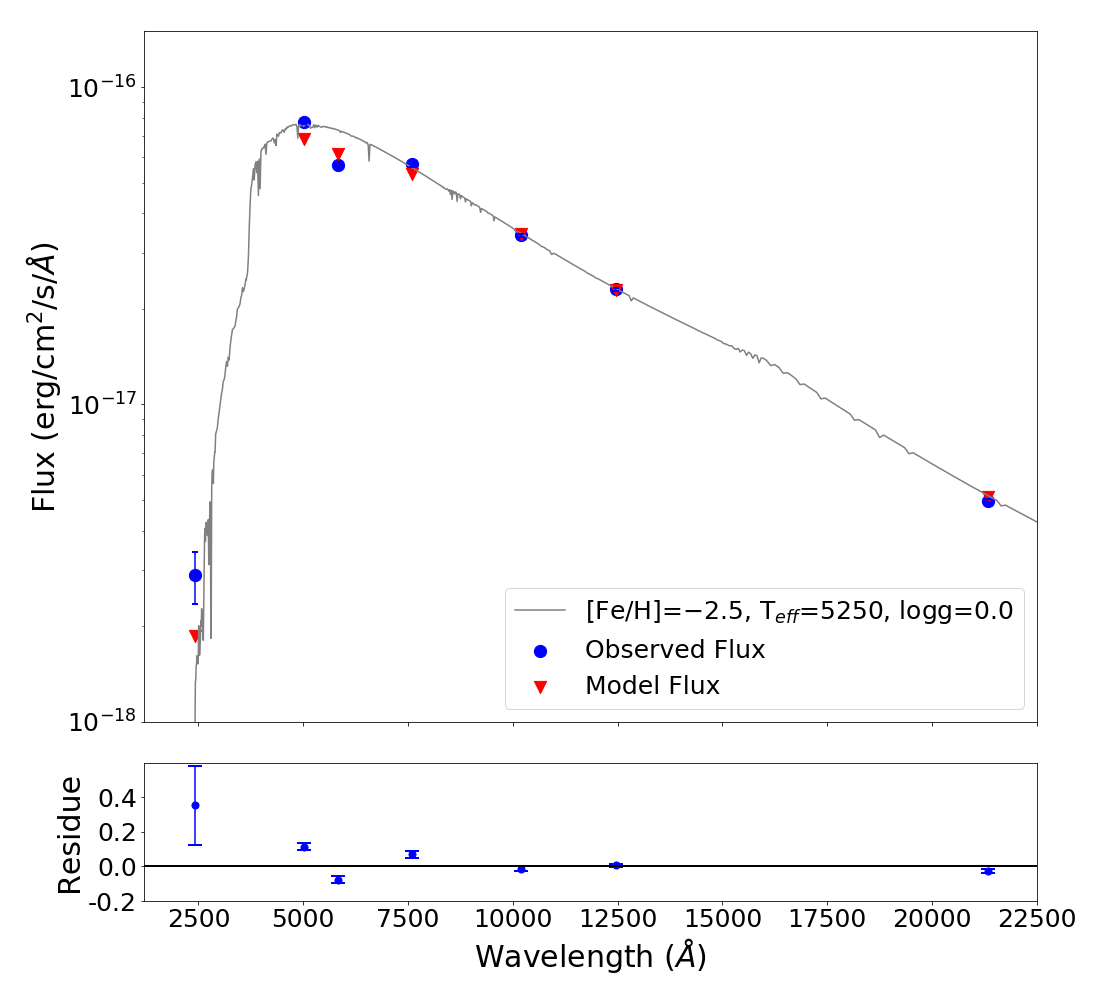}} \ 
{\includegraphics[width=0.74\columnwidth]{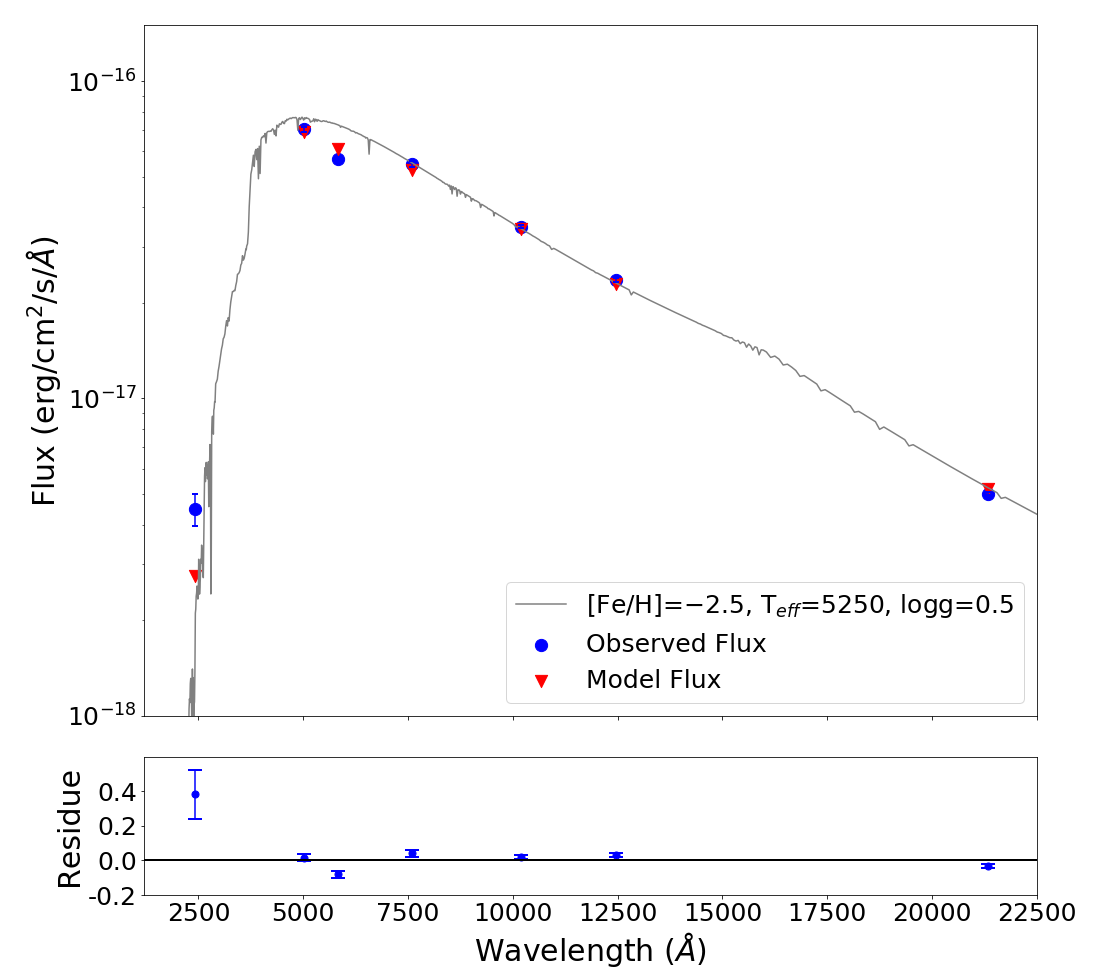}} {\includegraphics[width=0.74\columnwidth]{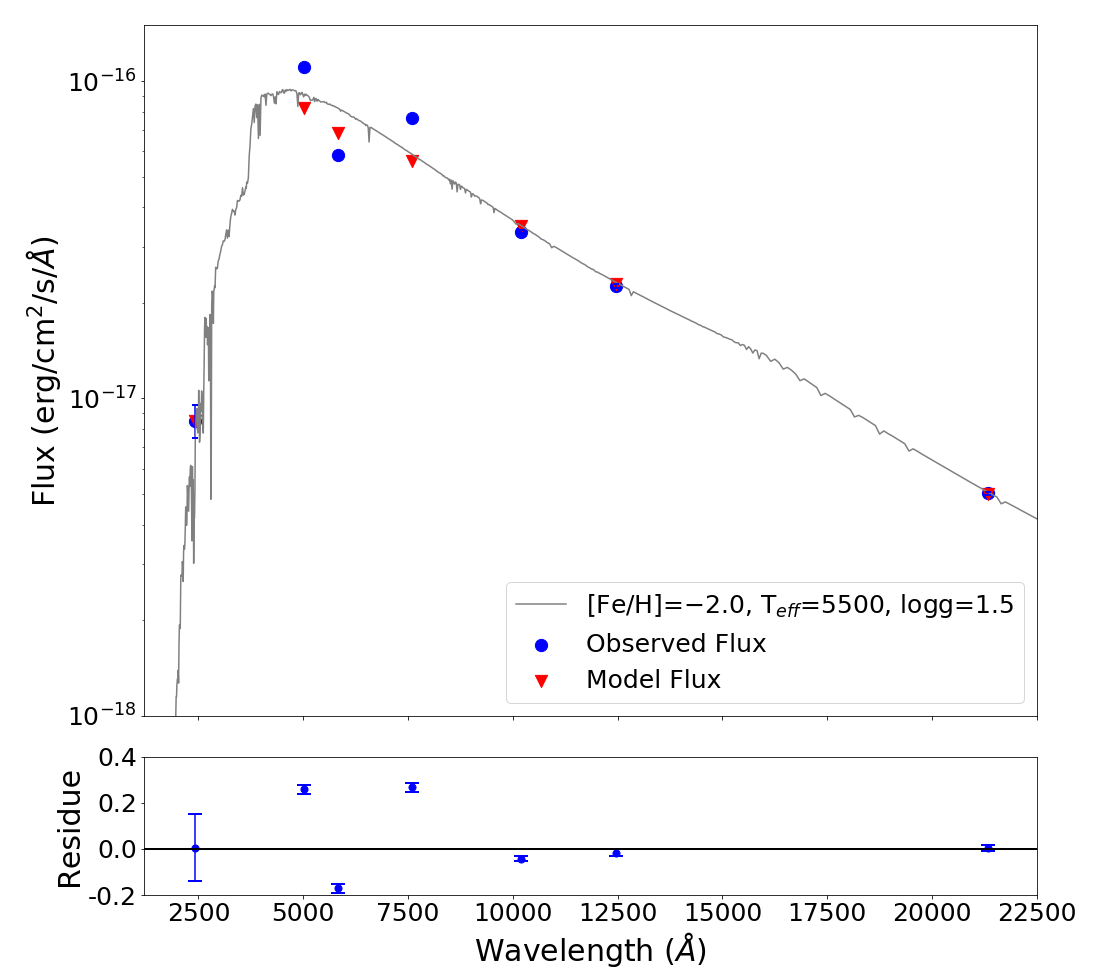}} 
\caption{The Figure represents the same as Fig. \ref{sed3}, but for another eight RC stars.}
\label{sed4}
\end{figure*}


\bsp	
\label{lastpage}
\end{document}